\documentclass[twocolumn, superscriptaddress, amsmath, amssymb, aps, physrev]{revtex4-2}
\usepackage{graphicx, xcolor, hyperref, dcolumn, bm}

\begin{document}

\title{Effects of cell-cell communication on bacterial chemotaxis}

\author{Soutick Saha}
\affiliation{Department of Physics and Astronomy, Purdue University, West Lafayette, IN 47907, USA}

\author{Sean Fancher}
\affiliation{Department of Physics and Astronomy, Purdue University, West Lafayette, IN 47907, USA}
\affiliation{Department of Biophysics, University of Michigan, Ann Arbor, Michigan 48109, USA}

\author{Andrew Mugler}
\email{Contact author: andrew.mugler@pitt.edu}
\affiliation{Department of Physics and Astronomy, Purdue University, West Lafayette, IN 47907, USA}
\affiliation{Department of Physics and Astronomy, University of Pittsburgh, Pittsburgh, Pennsylvania 15260, USA}

\begin{abstract}
    Bacteria track chemical gradients using a biased random walk, a process called chemotaxis. Experiments suggest that bacteria also communicate during this process. Using a mathematical model, we find that sufficiently strong communication succeeds in keeping a population of bacteria together but slows down chemotaxis. However, if the secretion of the communication molecule is coupled to the detection of the external chemoattractant, chemotaxis can be faster than without communication. Intriguingly, in this regime we predict that, even though blocking the communication receptors should slow down chemotaxis, partially blocking or underexpressing them should speed it up. Our work provides physical insights on how communication and chemotaxis are connected and may help explain why chemotaxing bacteria communicate.
\end{abstract}

\maketitle

\section{\label{sec:Introduction}Introduction}
Cells rarely exist in isolation, and communication has been shown to lie at the heart of efficient functioning in cells \cite{su2024cell}. The effects of communication can be observed in different biological systems, ranging from bacteria to mammalian cells \cite{fancher2017fundamental}. Cell-cell communication is central to bacterial quorum sensing \cite{mukherjee2019bacterial}, antibiotic resistance \cite{huang2020roles}, cancer progression \cite{dominiak2020communication}, and development \cite{gilbert2014developmental}. Understanding the role of communication in different biological processes is pivotal to understanding cellular behavior.

At the same time, most cells migrate. Cell migration is critical for development \cite{aman2010cell, cheung2025collective}, proper function of the immune system \cite{luster2005immune}, cancer metastasis \cite{stuelten2018cell, cheung2025collective}, and bacterial survival \cite{wadhwa2022bacterial}. Many single-celled organisms migrate to find more favorable environments. These migrations often take place in a biased way in response to chemical gradients, a process known as chemotaxis \cite{tu2013quantitative}.

Both communication and migration are ubiquitous in bacteria. While both capabilities have long been thought to be deeply connected \cite{jacob2004bacterial}, most of our knowledge of their interplay comes from a few key experimental examples. It was shown that on certain substrates, \textit{Escherichia coli} bacteria both secrete and chemotax toward a single chemical chemoattractant \cite{budrene1991complex}. Similarly, it was shown that starved \textit{E.\ coli} in confinement secrete and chemotax toward a particular amino acid \cite{park2003motion, park2003influence}. These studies demonstrate that bacteria not only follow attractants but can be their source.

Most recently, it was shown that \textit{E.\ coli} migrate faster toward a known chemoattractant compared to a strain where a receptor that has no role in sensing the chemoattractant was inactivated \cite{long2017cell}. This result suggests that the bacteria use the inactivated receptor to sense a secondary communication molecule that benefits chemotaxis. Here the roles are chemotaxis and communication are distinct but clearly connected. The precise nature of this connection remains unclear.

Motivated by these most recent experiments, here we ask: Can cell-cell communication speed up bacterial chemotaxis, and if so, is there an optimal communication strategy? To answer these questions we write down a continuum mathematical model for chemotaxis based on the Keller-Segel equations \cite{keller1971traveling}, and we incorporate cell-cell communication into this model. We analytically derive the chemotactic drift speed and cell dispersion under the approximation that the cell distribution is Gaussian, and we show that these results agree with numerical solution of the model.

We find that a co-attracting communication molecule secreted at a constant rate is always detrimental to chemotaxis because the molecule lags behind the cell population, decreasing the chemotactic drift. However, when the communication is adaptive, i.e., the secretion rate is coupled to the external chemoattractant concentration, there is a critical adaptive strength above which communication benefits chemotaxis. Our results are consistent with the experiments and suggest that the modules that govern communication and chemotaxis in cell signaling networks must be coupled for communication to be beneficial \cite{long2017cell}. Furthermore, our results lead to the surprising prediction that, while fully inactivating the communication receptors should slow down chemotaxis, partially inactivating them should speed it up.

\section{\label{sec:Results}Results}
\subsection{\label{sec:KS_eqs}Mathematical model}

\begin{figure}
    \centering
    \includegraphics[width=1\linewidth]{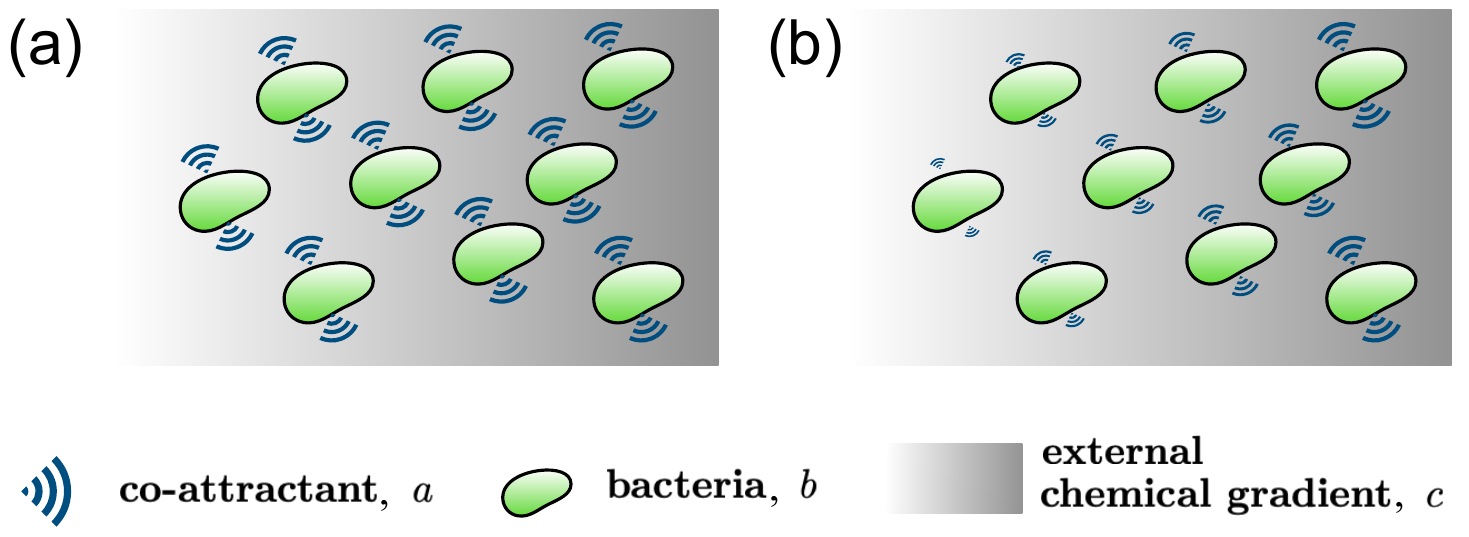}
    \caption{Schematic of chemotaxis with communication. A population of bacteria (green) migrates up the gradient of an external chemical (gray) and secretes a diffusing chemical (blue) to which it is also attracted. (a) Bacteria secrete the co-attractant at a constant rate. (b) The secretion rate increases with external chemical concentration, which we term adaptive communication.}
    \label{fig:chem_strategies}
\end{figure}

Population-level chemotaxis was first modeled by Keller and Segel in the context of slime mold \cite{keller1970initiation} and bacteria \cite{keller1971traveling}, and their framework has been widely used to investigate bacterial chemotaxis \cite{tindall2008overview}. Our model follows this framework and incorporates long-range cell-cell communication via a secreted, diffusing, co-attracting molecule \cite{salman2006solitary} [Fig. \ref{fig:chem_strategies}(a)]. We work in one spatial dimension for simplicity, but we expect our results to hold in higher dimensions as well (see Discussion and Appendix \ref{moments}).

Consider a bacterial population with density $b(x,t)$ and a co-attractant with concentration $a(x,t)$. The co-attractant is secreted with rate $\beta$ and degrades with rate $\nu$. The cells are exposed to an external chemical attractant with concentration $c(x)$. We assume that $c$ remains constant in time, which occurs if consumption does not significantly disrupt the profile, or if the attractant is rapidly replenished.

Studies have shown that \textit{E. coli} detects the fold change in attractant concentrations rather than the concentration itself \cite{lazova2011response}. Therefore, in our model, cells detect the fold change in the concentration of both the co-attractant and the external chemical. This leads to two partial differential equations for the spatiotemporal dynamics of the bacteria and co-attractant,
\begin{align}
\label{eq:KS1}
\partial_t b &= D_b\partial_x[\partial_x b - \gamma b\partial_x(\ln c)-\alpha b\partial_x(\ln a)], \\
\partial_t a &= D_a\partial_x^2 a + \beta b - \nu a
\label{eq:KS},
\end{align}
Here, $D_b$ and $D_a$ are the diffusion coefficients of the bacterial cells and the co-attractant molecules, respectively, while $\gamma$ and $\alpha$ are the strengths with which the cells detect the external chemical and the co-attractant, respectively, scaled by $D_b$. Describing the cells as a diffusing field coarse-grains over their run-and-tumble random walks, which is a hallmark of the Keller-Segel approach. The natural logs incorporate the fold-change detection, since responding to the derivative of the log of a quantity is equivalent to responding to its fold change. The second and third terms on the righthand side of Eq.\ \ref{eq:KS1} then constitute drift terms, pulling the bacterial population $b$ in the directions of higher fold change in the external attractant $c$ and co-attractant $a$, respectively.

For simplicity, we consider an exponential profile for the external chemical,
\begin{equation}
\label{c}
c(x) = c_0e^{gx},
\end{equation}
where $g^{-1}$ is the profile lengthscale, and $c_0$ is its concentration at the origin. In one dimension, an exponential profile is the steady state for a chemical diffusing from a point source with spontaneous decay \cite{wartlick2009morphogen}, and in higher dimensions this profile is enveloped by an exponential \cite{fancher2017fundamental}. Here, the convenience is that an exponential has a constant fold change, $\partial_x(\ln c) = g$.

We numerically solve Eqs.\ \ref{eq:KS1}-\ref{c}, initializing both $a$ and $b$ as Gaussian distributions at the origin, and using the following parameter values: $D_a = D_b = 400$ $\mu$m$^2$/s, $\beta = \nu = 10$/s, $\gamma = 1$, and $g^{-1} = 50$ $\mu$m, with $\alpha$ varied. We find that the results below are not qualitatively affected by changes to the initial conditions or to the parameter values, an observation that we later support with our analytic results. The code is freely available \cite{code}.

\subsection{\label{sec:non_adaptive_spread} Bacteria co-localize above a critical communication strength}

\begin{figure}
    \centering
    \includegraphics[width=\linewidth]{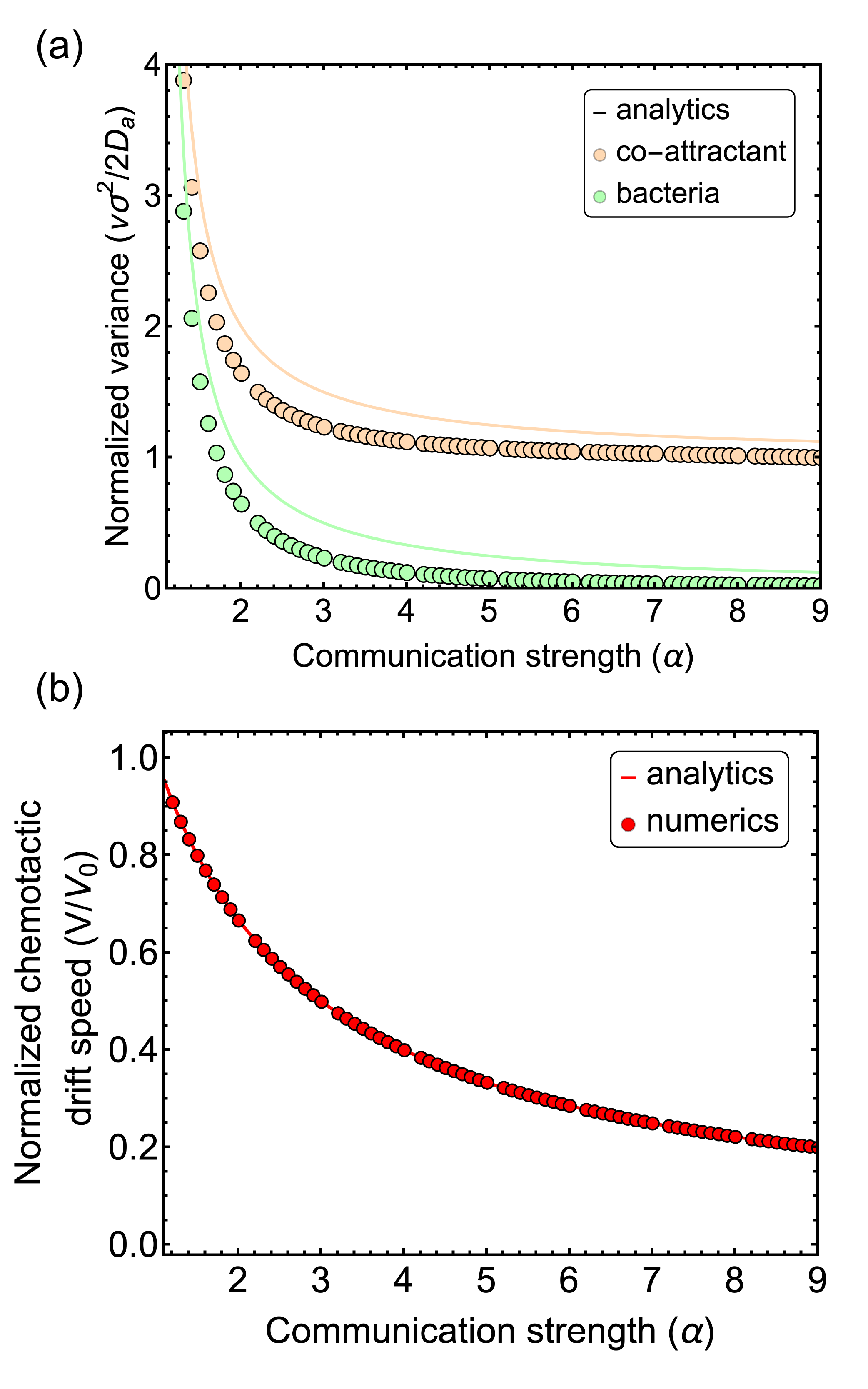}
    \caption{Effects of constant communication on chemotaxis. (a) Both the bacterial population and co-attractant profile maintain a finite width in steady state for $\alpha > 1$. Here there is no external attractant ($c=0$). (b) Constant communication slows down chemotaxis. Here $V_0 \equiv D_b\gamma g$. In both panels, circles are numerical results, and curves are the analytic expressions (Eqs.\ \ref{var} and \ref{V}, respectively).}
    \label{fig:non_adaptive}
\end{figure}

First we investigate the effect of communication on bacterial cells in the absence of an external chemical gradient ($c = 0$). In this scenario we observe from our numerical solutions that both the bacterial and the co-attractant densities remain localized, even after long times, if the communication strength is above a critical value $\alpha\approx1$. Below this value, both densities disperse at long times. This behavior can be seen by plotting from the numerics the variance of the bacterial and co-attractant distributions after they have converged at long times, as a function of the communication strength $\alpha$, as shown in Fig.\ \ref{fig:non_adaptive}(a) (green and orange circles). We see that as $\alpha$ decreases, both variances increase, until eventually diverging near $\alpha\approx1$.

Qualitatively, the reason for finite variances at long times is that communication promotes co-attraction, whereas diffusion promotes spread. The balance between these effects leads to finite variances for both bacteria and co-attractant in steady state. It makes sense that the variances decrease with $\alpha$ because increasing communication strengthens the attraction effect. It also makes sense that the variance of the co-attractant is larger than that of the bacteria because whereas both bacteria and co-attractant diffuse, only the bacteria chemotax. Evidently, below $\alpha\approx1$, communication is no longer sufficient to balance diffusion, and the distributions spread indefinitely, causing the variance to diverge.

To understand these effects quantitatively, we investigate Eqs.\ \ref{eq:KS1} and \ref{eq:KS} analytically. Following previous work \cite{young2019interactions}, we approximate the distributions of bacteria and co-attractant as Gaussian,
\begin{equation}
\label{Gaussian}
b = \frac{Be^{-(x-\mu_b)^2/(2\sigma_b^2)}}{\sqrt{2\pi\sigma_b^2}}, \quad
a = \frac{Ae^{-(x-\mu_a)^2/(2\sigma_a^2)}}{\sqrt{2\pi\sigma_a^2}},
\end{equation}
with time-dependent amplitudes $B(t)$ and $A(t)$, mean positions $\mu_b(t)$ and $\mu_a(t)$, and spatial variances $\sigma_b^2(t)$ and $\sigma_a^2(t)$. The dynamics of these parameters come from integrating Eqs.\ \ref{eq:KS1} and \ref{eq:KS} (with $c=0$) over all space against $1$, $x$, or $x^2$. Noting from Eq.\ \ref{Gaussian} that $\partial_x(\ln a) = (\mu_a-x)/\sigma_a^2$, this procedure gives (see Appendix \ref{moments})
\begin{align}
\label{noc1}
\partial_t B =&\ 0, \\
\label{noc2}
\partial_t A =&\ \beta B - \nu A, \\
\label{noc3}
\partial_t \mu_b =&\ D_b\alpha(\mu_a-\mu_b)/\sigma_a^2, \\
\label{noc4}
\partial_t \mu_a =&\ \beta B(\mu_b-\mu_a)/A, \\
\label{noc5}
\partial_t (\sigma_b^2) =&\ 2D_b(1-\alpha\sigma_b^2/\sigma_a^2), \\
\label{noc6}
\partial_t (\sigma_a^2) =&\ 2D_a
    + \beta B[(\mu_b-\mu_a)^2 + \sigma_b^2-\sigma_a^2]/A.
\end{align}
In deriving each of these equations, we exploit the fact that a Gaussian and its derivative vanish as $x\to\pm\infty$ more strongly than a polynomial, to neglect the boundary terms that arise from integration by parts.

The steady state of Eqs.\ \ref{noc1}-\ref{noc6} is: $B =$ constant, which makes sense because no bacteria are gained or lost in our model; $A = \beta B/\nu$, which shows that the total co-attractant amount in steady state balances bacteria-dependent production against degradation; $\mu_a = \mu_b = $ constant, which must be true by symmetry (with $c=0$); and
\begin{equation}
\label{var}
\sigma_a^2 = \alpha\sigma_b^2 = \frac{2\alpha D_a}{\nu(\alpha-1)}
\qquad (\alpha>1).
\end{equation}
Equation \ref{var} confirms that both variances diverge as $\alpha$ approaches the critical value $1$ from the right. It also confirms, since $\alpha>1$, that the variance of the co-attractant is larger than that of the bacteria. Equation \ref{var} is compared with the numerical results in Fig.\ \ref{fig:non_adaptive}(a) (green and orange curves), and we see good agreement, including the divergent behavior, with small discrepancy due to the Gaussian approximation.

\subsection{\label{sec:non_adaptive_chemotaxis} Constant communication slows down chemotaxis}
Next we consider the external attractant with an exponential profile (Eq.\ \ref{c}). For subcritical communication ($\alpha<1$),
we observe from our numerical solutions of Eqs.\ \ref{eq:KS1}-\ref{eq:KS} that the chemotactic drift speed of the bacteria does not change with the communication strength $\alpha$. This makes sense because in this regime the bacteria spread indefinitely and therefore chemotax as single cells; communication plays no role at long times.
However, for $\alpha>1$, we observe
that the chemotactic drift speed changes with $\alpha$. Figure \ref{fig:non_adaptive}(b) (red circles) shows the drift speed as a function of $\alpha$, normalized by its value when $\alpha=0$. We see that this quantity is always less than one, and it decreases with $\alpha$. Hence, in this model, when communication molecules are secreted at a constant rate, communication always slows down chemotaxis.

To understand this result quantitatively, we again turn to the Gaussian approximation of Eq.\ \ref{Gaussian}. Recalling that $\partial_x(\ln c) = g$, and performing the same integration procedure as before (see Appendix \ref{moments}), we find that Eqs.\ \ref{noc1}, \ref{noc2}, and \ref{noc4}-\ref{noc6} are unchanged, while Eq.\ \ref{noc3} becomes
\begin{equation}
\label{c3}
\partial_t \mu_b = D_b\gamma g + D_b\alpha(\mu_a-\mu_b)/\sigma_a^2.
\end{equation}
The first term on the righthand side is new: it is the drift velocity in the absence of communication, which we define as $V_0\equiv D_b\gamma g$.

Due to the drift, the mean positions of the bacteria and co-attractant no longer reach steady state. However, because the bacteria both secrete and follow the co-attractant, we make the assumption that the drift speeds of bacteria and co-attractant are equal at long times. Calling this speed $V\equiv\partial_t\mu_b = \partial_t\mu_a$, and again assuming the co-attractant reaches its steady total amount $A=\beta B/\nu$, Eqs.\ \ref{noc4} and \ref{c3} are solved to give
\begin{equation}
\label{Vtemp}
V = \nu\Delta = \frac{\nu V_0}{\nu+D_b\alpha/\sigma_a^2},
\end{equation}
where we have defined $\Delta\equiv\mu_b-\mu_a$ as the separation between the two mean positions. If we again assume that the variances of bacteria and co-attractant approach constant values at long times, even as the distributions are drifting, then Eqs.\ \ref{noc5} and \ref{noc6} in steady state give
\begin{equation}
\label{var2}
\sigma_a^2 = \alpha\sigma_b^2 = \frac{\alpha(2D_a+\nu\Delta^2)}{\nu(\alpha-1)}
\qquad (\alpha>1).
\end{equation}
Eq.\ \ref{var2} is Eq.\ \ref{var} with an extra term in the numerator that arises because symmetry is not preserved in the presence of the external gradient ($\Delta\ne0$). We find that this term can be neglected if communication is sufficiently fast, $D_a\nu\gg V_0^2$, which is justified for bacteria (see Appendix \ref{fast}). In this case, Eq.\ \ref{Vtemp} becomes
\begin{equation}
\label{V}
V = \nu\Delta = \frac{V_0}{1+(\alpha-1)D_b/(2D_a)}
\qquad (\alpha>1).
\end{equation}
We see that the separation, and therefore the common speed of the distributions, attains a constant value. As anticipated, the speed approaches $V_0$ as $\alpha\to1$, and it decreases as $\alpha$ increases. Equation \ref{V} is compared with the numerical results in Fig.\ \ref{fig:non_adaptive}(b) (red curve), and we see excellent agreement.

Furthermore, we see from Eq.\ \ref{V} that $\Delta$ is positive, meaning that $\mu_b>\mu_a$: the bacteria lead the co-attractant. This makes sense because by the time the co-attractant has diffused and degraded, the bacteria have moved to the right, up the external gradient. It also helps explain why communication slows down chemotaxis under this protocol: the bacteria are attracted to both the external attractant and the co-attractant; the external attractant pulls them forward while the co-attractant pulls them back, reducing their speed.

\subsection{\label{sec:adaptive_chemotaxis} Adaptive communication can speed up chemotaxis}

If we propose that detection of the external attractant and secretion of the co-attractant are coupled \cite{long2017cell}, can communication speed up chemotaxis? To address this question, we allow the secretion rate $\beta$ to depend on the external concentration $c$ [Fig.\ \ref{fig:chem_strategies}(b)]. Specifically, we write
\begin{equation}
\label{beta}
\beta = \beta_0 + \beta_1(\ln c - \langle\ln c\rangle),
\end{equation}
where $\beta_0$ and $\beta_1$ are constants, and the average is taken over the normalized distribution of bacteria, $b(x,t)/B(t)$. Subtracting the average makes secretion adaptive: bacteria that detect a (logarithmic) concentration larger than the population average will secrete more, and vice versa. Therefore, we call this case adaptive communication. We do not specify the mechanism by which a cell acquires knowledge of the population average, but we note that adaptive sensing, which is ubiquitous in single cells \cite{tu2013quantitative, tang2014evolutionarily, takeda2012incoherent}, has been theoretically proposed for cell populations \cite{mugler2016limits, camley2016collective} and supported by data from mammalian cell clusters \cite{ellison2016cell}. In sparse cell collectives like chemotaxing bacteria, population information would need to be mediated by a long-range signal \cite{fancher2017fundamental}, and it has been shown theoretically that the co-attractant itself is sufficient for this task \cite{gonzalez2023collective}.

\begin{figure}
    \centering
    \includegraphics[width=.95\linewidth]{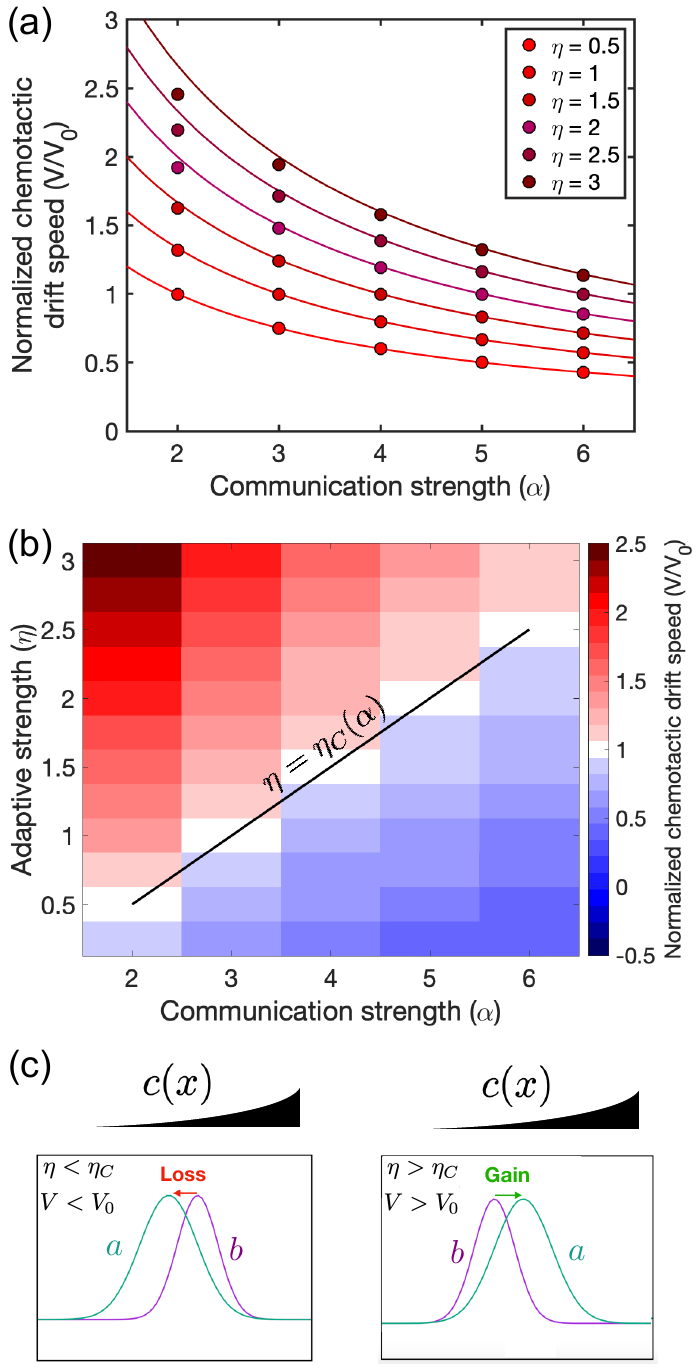}
    \caption{Effects of adaptive communication on chemotaxis. (a) Adaptation (increasing $\eta$) increases chemotaxis speed. Circles are numerical results, and curves are Eq.\ \ref{Va}. (b) Communication speeds up chemotaxis for $\eta > \eta_c$. Color map is numerical results, and black line is Eq.\ \ref{etac}. (c) Schematic of the mechanism: for $\eta < \eta_c$, the co-attractant lags behind the bacteria, slowing down chemotaxis; for $\eta > \eta_c$, the co-attractant leads the bacteria, speeding up chemotaxis.}
    \label{fig:adaptive}
\end{figure}

According to Eq.\ \ref{beta}, communication is more adaptive for larger $\beta_1$ or smaller $\beta_0$. Therefore, we define a dimensionless adaptation strength,
\begin{equation}
\label{eta}
\eta \equiv \frac{\beta_1}{\gamma\beta_0},
\end{equation}
where we have included a factor of $\gamma$ for later convenience. Numerical solutions of Eqs.\ \ref{eq:KS1}-\ref{c} with $\beta$ as in Eq.\ \ref{beta} are shown in Fig.\ \ref{fig:adaptive}(a) (circles) for various values of $\eta$ (colors). We see that, while the drift speed still decreases with communication strength $\alpha$, it increases with adaptation strength $\eta$. In fact, we see that for sufficiently large $\eta$, we can have $V/V_0>1$, meaning that the drift speed can be larger than its value in the absence of communication. Thus, in this regime, adaptive communication speeds up chemotaxis.

To understand this result quantitatively, we turn a final time to the Gaussian approximation of Eq.\ \ref{Gaussian}. Noting from Eq.\ \ref{c} that $\ln c - \langle\ln c\rangle = g(x-\mu_b)$, and performing the same integration procedure as before (see Appendix \ref{moments}), we find that Eqs.\ \ref{noc1}, \ref{noc2}, \ref{noc5}, and \ref{c3} are unchanged (with $\beta\to\beta_0$), while Eqs.\ \ref{noc4} and \ref{noc6} become
\begin{align}
\label{c4}
\partial_t \mu_a =&\ \beta_0 B(\Delta+\eta V_0\sigma_b^2/D_b)/A, \\
\label{c6}
\partial_t (\sigma_a^2) =&\ 2D_a
    + \beta_0 B[\Delta^2 + \sigma_b^2-\sigma_a^2
    + 2\eta V_0\sigma_b^2\Delta/D_b]/A.
\end{align}
The term involving $\eta$ in each equation is new, arising from the adaptive contribution to the secretion of the co-attractant.
Solving Eqs.\ \ref{noc5} and \ref{c6} in steady state now gives
\begin{equation}
\label{var3}
\sigma_a^2 = \alpha\sigma_b^2
= \frac{\alpha(2D_a+\nu\Delta^2)}{\nu(\alpha-1-2\eta V_0\Delta/D_b)}
\qquad (\alpha>1),
\end{equation}
where the term involving $\eta$ is new. However, we find that this term, like the second term in the numerator, can be neglected in the fast communication limit $D_a\nu\gg V_0^2$ (see Appendix \ref{fast}), reducing Eq.\ \ref{var3} to Eq.\ \ref{var}.
We can then, as before, assume $A=\beta_0 B/\nu$ and solve Eqs.\ \ref{c3} and \ref{c4} for the common drift speed $V\equiv\partial_t\mu_b = \partial_t\mu_a$ and the mean separation $\Delta$.

The speed is
\begin{equation}
\label{Va}
V = \frac{V_0(1+\eta)}{1+(\alpha-1)D_b/(2D_a)}
\qquad (\alpha>1).
\end{equation}
Comparing with Eq.\ \ref{V}, we see that adaptive communication simply multiplies the speed by a factor $1+\eta$. This is why the numerical results in Fig.\ \ref{fig:adaptive}(a) appear to simply scale with $\eta$, while the $\alpha$ dependence remains unchanged. Indeed, Eq.\ \ref{Va} is plotted in Fig.\ \ref{fig:adaptive}(a) (curves), and we see good agreement with the numerical results.

The form of Eq.\ \ref{Va} suggests that we define a critical adaptation strength,
\begin{equation}
\label{etac}
\eta_c \equiv \frac{(\alpha-1)D_b}{2D_a},
\end{equation}
in terms of which Eq.\ \ref{Va} reads
\begin{equation}
\label{Va2}
\frac{V}{V_0} = \frac{1+\eta}{1+\eta_c}.
\end{equation}
Equation \ref{Va2} makes clear that when $\eta > \eta_c$, we have $V>V_0$, and vice versa. Therefore, $\eta_c$ is the minimum adaptation strength for which communication speeds up chemotaxis. We have checked that it is possible to have $\eta>\eta_c$ without $\beta$ (Eq.\ \ref{beta}) becoming negative within the bacterial population, so long as the same fast-communication limit $D_a\nu\gg V_0^2$ holds (Appendix \ref{positivity}).

The prediction that $V>V_0$ when $\eta > \eta_c$ is tested in Fig.\ \ref{fig:adaptive}(b). There, we plot $V/V_0$ from numerical results as a function of $\alpha$ and $\eta$ (color map), and we overlay Eq.\ \ref{etac} (line). We see that, as expected, communication slows down chemotaxis in the regime $\eta<\eta_c$ (blue), and communication speeds up chemotaxis in the regime $\eta>\eta_c$ (red).

Meanwhile, the mean separation is
\begin{equation}
\label{Delta}
\Delta = \frac{V_0}{\nu}\left[\frac{\eta_c-\eta}{\eta_c(1+\eta_c)}\right].
\end{equation}
Equation \ref{Delta} makes clear why communication is beneficial for $\eta>\eta_c$: this condition makes $\Delta < 0$. Because $\Delta = \mu_b-\mu_a$, this means that $\mu_a>\mu_b$: the co-attractant leads the bacteria. The reason is that cells near the front secrete more than cells near the back, due to the adaptive nature of the secretion. The outcome is that there is no longer a conflict between following the external attractant and following the co-attractant; both pull the bacteria up the external gradient.

In summary, below a critical adaptation strength, the co-attractant lags behind the bacteria [Fig.\ \ref{fig:adaptive}(c), left]. This results in a slowdown of chemotaxis. Above the critical adaptation strength, the co-attractant leads the bacteria [Fig.\ \ref{fig:adaptive}(c), right]. This results in a speedup of chemotaxis.

\section{\label{sec:Discussion}Discussion}

Chemotaxis and cell-cell communication are separately well-studied, but their interplay is still poorly understood. Motivated by experiments suggesting that bacteria communicate during chemotaxis \cite{long2017cell}, we have investigated a continuum model of this interplay. We have found that communicating via a co-attracting chemical can co-localize bacteria but slows their chemotactic drift up an external chemical gradient. Yet, coupling the co-attractant secretion to detection of the external chemical boosts the chemotactic drift and can make chemotaxis faster than without communication. Our work provides analytic descriptions of these effects and reveals the mechanism: chemotaxis is slowed when the co-attractant lags behind the bacteria, and it is sped up when the co-attractant leads the bacteria. Coupling secretion to external detection, which we call adaptive communication, leads to the latter case because bacteria at the front of the pack secrete more co-attractant.

Co-localization of chemotaxing bacteria has been observed previously \cite{adler1966chemotaxis, phan2024direct} and has been attributed to consumption: by consuming the external attractant, bacteria steepen its gradient, and this enhanced attraction overpowers the tendency of run-and-tumble diffusion to disperse the bacteria \cite{keller1971traveling, narla2021traveling, phan2024direct}. Here, we demonstrate a distinct mechanism of co-localization arising from co-attractant secretion, applicable in cases where an external attractant is minimally consumed or is rapidly replenished. In principle, these two mechanisms could act simultaneously \cite{salman2006solitary}, and their interplay is an interesting direction for future work.

The experimental results suggesting that communication aids chemotaxis were supported by agent-based simulations in which bacteria secrete a co-attractant only when moving up the external gradient rather than indiscriminately \cite{long2017cell}. Our results are consistent with this finding: communication benefits chemotaxis only when up-gradient bacteria secrete more than down-gradient bacteria (adaptive communication) and not when secretion is indiscriminate. Our work provides a theoretical basis for this requirement, which complements and adds physical intuition to agent-based simulations. It also reveals the basic mechanism: the secreted chemical must lead the bacteria, on average, not lag behind them.

To our knowledge, the putative co-attractant molecule in {\it E.\ coli} has not been identified. However, if it were, our work makes predictions for experiments that would perturb it. Specifically, we have found that secretion of this molecule only benefits chemotaxis when coupled to detection of the external attractant. An experiment that knocks out the endogenous expression of this molecule and replaces it with externally inducible expression would remove this coupling. Our results predict that this perturbation would remove the benefit to chemotaxis. In fact, the chemotaxis speed should be slower than without communication at all [Fig.\ \ref{fig:non_adaptive}(b)]. More generally, our results suggest that in any cell type, the signaling modules that govern communication and chemotaxis must be tightly coupled for communication to be beneficial.

Our parameter $\alpha$ models the cells' sensitivity to the co-attractant. Therefore, $\alpha$ should correlate with the abundance or activity of the co-attractant receptors. The experiments that blocked these putative receptors observed a slowdown in chemotaxis \cite{long2017cell}. This observation is consistent with our model: in the regime where communication is beneficial ($\eta > \eta_c$) we can have $V > V_0$, and taking $\alpha\to0$ results in $V\to V_0$: a chemotaxis slowdown. But this regime has an important feature, seen in Fig.\ \ref{fig:adaptive}(a): if $\alpha$ simply decreases, rather than vanishing completely, then the chemotaxis speed actually increases. This holds true until $\alpha$ is lowered past its critical value $\alpha=1$, when the speed discontinuously drops to $V_0$ because the cells no longer hold together at long times.

The above observation leads to a surprising prediction: inducing a small decrease in the receptor sensitivity, for example by underexpressing the receptors or exposing them to an external supply of the co-attractant molecule, would speed cells up, even though blocking receptors completely would slow cells down. Conversely, overexpressing the receptors would, for the same reason, slow cells down. Essentially, our model predicts that cells are on one side of a discontinuous optimum, so that a local decrease and a global decrease in receptor activity have opposite effects.

We have worked in one spatial dimension for simplicity, but we expect our results to hold in higher dimensions. In fact, in Appendix \ref{moments}, we show that one of our main results, Eq.\ \ref{var}, holds in any dimension. This means that in the absence of the external chemical, bacteria co-localize above the critical communication strength $\alpha = 1$ in any spatial dimension. Extending our results in the presence of the external chemical is more difficult because the chemical gradient breaks spherical symmetry in higher dimensions. Nevertheless, because Eq.\ \ref{var} remained a good approximation for our one-dimensional results in the presence of the external chemical, we anticipate that these results will hold quantitatively or qualitatively in higher dimensions as well.

In this work we focused on the long-time behavior of collective chemotaxis, obtaining results for steady-state abundances, variances, and drift speeds. Future work could investigate transient effects, especially in the subcritical communication regime ($\alpha < 1$) where the variance grows without bound but communication could affect chemotaxis speed at short times. This regime could also be relevant for the experiments, where chemotaxis is observed in finite chambers over finite times \cite{long2017cell}. Future work could also investigate how our results are affected by cell-to-cell heterogeneity, which can be pronounced in chemotaxing bacterial populations \cite{waite2018behavioral}.

Although our work is motivated by experiments on bacteria, our results are generic. The mechanism we describe requires gradient sensing coupled to molecule secretion and uptake, which are ubiquitous features across cell types. Collective chemotaxis is observed in non-bacterial systems \cite{camley2018collective}, and therefore we anticipate that our findings will find broad application.

\acknowledgements
We thank Hanna Salman for a critical reading of the manuscript. This work was supported by National Science Foundation Grant No.\ PHY-2118561.

\appendix
\section{Moment dynamics}
\label{moments}
Here we derive the dynamics of the spatial moments of the bacteria and co-attractant distributions. We also show that Eq.\ \ref{var} for the variance in steady state holds in any spatial dimension.

The moments of Eq.\ \ref{Gaussian} are defined
\begin{align}
\label{mom1}
B &= \int_{-\infty}^{\infty}dx\ b, \\
\label{mom2}
A &= \int_{-\infty}^{\infty}dx\ a, \\
\label{mom3}
\mu_b &= \frac{1}{B}\int_{-\infty}^{\infty}dx\ xb, \\
\label{mom4}
\mu_a &= \frac{1}{A}\int_{-\infty}^{\infty}dx\ xa, \\
\label{mom5}
v_b + \mu_b^2 &= \frac{1}{B}\int_{-\infty}^{\infty}dx\ x^2b, \\
\label{mom6}
v_a + \mu_a^2 &= \frac{1}{A}\int_{-\infty}^{\infty}dx\ x^2a,
\end{align}
where we have written the variances as $\sigma_b^2\equiv v_b$ and $\sigma_a^2\equiv v_a$ for brevity.
Isolating the integrals and differentiating with respect to time gives
\begin{align}
\label{dot1}
\int_{-\infty}^{\infty}dx\ \dot b &= \dot B, \\
\label{dot2}
\int_{-\infty}^{\infty}dx\ \dot a &= \dot A, \\
\label{dot3}
\int_{-\infty}^{\infty}dx\ x\dot b &= \partial_t(B\mu_b) = \dot B\mu_b + B\dot \mu_b, \\
\label{dot4}
\int_{-\infty}^{\infty}dx\ x\dot a &= \partial_t(A\mu_a) = \dot A\mu_a + A\dot \mu_a, \\
\label{dot5}
\int_{-\infty}^{\infty}dx\ x^2\dot b &= \partial_t[B(v_b+\mu_b^2)] \nonumber \\
&= \dot B(v_b+\mu_b^2) + B(\dot v_b + 2\mu_b\dot \mu_b), \\
\label{dot6}
\int_{-\infty}^{\infty}dx\ x^2\dot a &= \partial_t[A(v_a+\mu_a^2)] \nonumber \\
&= \dot A(v_a+\mu_a^2) + A(\dot v_a + 2\mu_a\dot \mu_a),
\end{align}
where we use dot for the time derivative when convenient. We will use Eqs.\ \ref{mom1}-\ref{dot6} in the derivations.

\subsection{No external attractant (Eqs.\ \ref{noc1}-\ref{noc6})}

Noting from Eq.\ \ref{Gaussian} that $\partial_x(\ln a) = (\mu_a-x)/v_a$, Eqs.\ \ref{eq:KS1} and \ref{eq:KS} with $c=0$ read
\begin{align}
\label{bdot1}
\dot b &= D_b[b'-\alpha b(\mu_a-x)/v_a]', \\
\label{adot1}
\dot a &= D_a a'' + \beta b - \nu a,
\end{align}
where we use prime for the spatial derivative when convenient.
Integrating Eq.\ \ref{bdot1} over $x$ gives
\begin{equation}
\label{Bdot1}
\dot B = D_b\int_{-\infty}^{\infty}dx\ [b'-\alpha b(\mu_a-x)/v_a]'.
\end{equation}
Eq.\ \ref{Bdot1} is an integral of a derivative, so we evaluate the term in brackets at the boundaries. Because both $b'$ and $b$ for a Gaussian vanish as $x\to\pm\infty$ more strongly than a polynomial, this term is zero. Thus,
\begin{equation}
\label{noc1A}
\dot B=0
\end{equation}
as in Eq.\ \ref{noc1}.

Integrating Eq.\ \ref{adot1} over $x$ gives
\begin{equation}
\dot A = D_a\int_{-\infty}^{\infty}dx\ a'' + \beta B - \nu A.
\end{equation}
Again, $a'$ vanishes at the boundaries, leaving
\begin{equation}
\label{noc2A}
\dot A = \beta B - \nu A   
\end{equation}
as in Eq.\ \ref{noc2}.

Multiplying Eq.\ \ref{bdot1} by $x$ and integrating over $x$ gives
\begin{equation}
\label{mubdot1}
\mu_b\dot B + \dot \mu_b B
= D_b\int_{-\infty}^{\infty}dx\ x[b'-\alpha b(\mu_a-x)/v_a]',
\end{equation}
where we have used Eq.\ \ref{dot3} on the lefthand side. Because $\dot B = 0$ (Eq.\ \ref{noc1A}), the first term vanishes. On the righthand side, we integrate by parts, giving
\begin{equation}
\label{mubdot2}
\dot \mu_b B
= - D_b\int_{-\infty}^{\infty}dx\ [b'-\alpha b(\mu_a-x)/v_a],
\end{equation}
where we ignore the boundary term since $b'$ and $b$ vanish strongly there. The first term is a derivative of an integral and vanishes at the boundaries, while the second term integrates to $B$ (Eq.\ \ref{mom1}) and $B\mu_b$ (Eq.\ \ref{mom3}). The $B$ cancels on both sides, leaving
\begin{equation}
\label{noc3A}
\dot \mu_b = D_b\alpha(\mu_a-\mu_b)/v_a
\end{equation}
as in Eq.\ \ref{noc3}.

Multiplying Eq.\ \ref{adot1} by $x$ and integrating over $x$ gives
\begin{equation}
\mu_a\dot A + \dot \mu_a A
= D_a\int_{-\infty}^{\infty}dx\ xa'' + \beta B\mu_b - \nu A\mu_a,
\end{equation}
where we have used Eqs.\ \ref{mom3}, \ref{mom4}, and \ref{dot4}. The first term on the righthand side is an integral of a derivative even after integrating by parts, and vanishes. Inserting Eq.\ \ref{noc2A} for $\dot A$ and simplifying gives
\begin{equation}
\label{noc4A}
\dot \mu_a = \beta B(\mu_b-\mu_a)/A
\end{equation}
as in Eq.\ \ref{noc4}.

Multiplying Eq.\ \ref{bdot1} by $x^2$ and integrating over $x$ gives
\begin{equation}
\label{noc5temp}
B(\dot v_b + 2\mu_b\dot \mu_b)
= D_b\int_{-\infty}^{\infty}dx\ x^2[b'-\alpha b(\mu_a-x)/v_a]',
\end{equation}
where we have used Eq.\ \ref{dot3}  with $\dot B = 0$ (Eq.\ \ref{noc1A}) on the lefthand side. Integrating by parts gives
\begin{equation}
B(\dot v_b + 2\mu_b\dot \mu_b)
= -2D_b\int_{-\infty}^{\infty}dx\ x[b'-\alpha b(\mu_a-x)/v_a].
\end{equation}
Integrating the first term by parts again gives $2D_bB$, while the second term integrates to $B\mu_b$ (Eq.\ \ref{mom3}) and $B(v_b+\mu_b^2)$ (Eq.\ \ref{mom5}). Inserting Eq.\ \ref{noc3A} for $\dot \mu_b$ and simplifying gives
\begin{equation}
\label{noc5A}
\dot v_b = 2D_b(1-\alpha v_b/v_a)
\end{equation}
as in Eq.\ \ref{noc5}.

Multiplying Eq.\ \ref{adot1} by $x^2$ and integrating over $x$ gives
\begin{align}
&\dot A(v_a+\mu_a^2) + A(\dot v_a + 2\mu_a\dot \mu_a) \nonumber \\
\label{noc6temp}
&= D_a\int_{-\infty}^{\infty}dx\ x^2a''
+ \beta B(v_b + \mu_b^2) - \nu A(v_a + \mu_a^2),
\end{align}
where we have used Eqs.\ \ref{mom5}, \ref{mom6}, and \ref{dot6}. Integrating the first term on the righthand side by parts twice gives $2D_aA$. Inserting Eq.\ \ref{noc2A} for $\dot A$ and Eq.\ \ref{noc4A} for $\dot \mu_a$ and simplifying gives
\begin{equation}
\label{noc6A}
\dot v_a = 2D_a + \beta B[(\mu_b-\mu_a)^2+v_b-v_a]/A
\end{equation}
as in Eq.\ \ref{noc6}.

\subsection{External attractant (Eq.\ \ref{c3})}

With $(\ln c)' = g$, Eqs.\ \ref{eq:KS1} and \ref{eq:KS} read
\begin{align}
\label{bdot2}
\dot b &= D_b[b'-\gamma bg - \alpha b(\mu_a-x)/v_a]', \\
\label{adot2}
\dot a &= D_a a'' + \beta b - \nu a.
\end{align}
Comparing to Eqs.\ \ref{bdot1} and \ref{adot1}, we see that Eq.\ \ref{bdot2} has an added term, while Eq.\ \ref{adot2} is unchanged. Because Eq.\ \ref{adot2} is unchanged, the equations for $\dot A$, $\dot \mu_a$, and $\dot v_a$ are unchanged.

Eq.\ \ref{bdot2} is still a total derivative, so integrating it gives $\dot B = 0$ as in Eq.\ \ref{noc1A}. But integrating Eq.\ \ref{bdot2} against $x$ gives a new term $-D_b\gamma g\int dx\ xb'$ on the righthand side. Integrating by parts makes it $D_b\gamma gB$. Thus, Eq.\ \ref{noc3A} becomes
\begin{equation}
\label{c3A}
\dot \mu_b = D_b\gamma g + D_b\alpha(\mu_a-\mu_b)/v_a.
\end{equation}
Similarly, integrating Eq.\ \ref{bdot2} against $x^2$ gives a new term $-D_b\gamma g\int dx\ x^2b'$ on the righthand side, and integrating by parts makes it $2D_b\gamma gB\mu_b$. But this term is canceled by the new $D_b\gamma g$ term in $\dot \mu_b$, once $\dot \mu_b$ is inserted into the lefthand side (see Eq.\ \ref{noc5temp}). This cancellation leaves Eq.\ \ref{noc5A} unchanged.

In summary, the only change is Eq.\ \ref{c3A}, as in Eq.\ \ref{c3}.

\subsection{Adaptive communication (Eqs.\ \ref{c4} and \ref{c6})}

With $\beta = \beta_0 + \beta_1g(x-\mu_b)$, Eqs.\ \ref{eq:KS1} and \ref{eq:KS} read
\begin{align}
\label{bdot3}
\dot b &= D_b[b'-\gamma bg - \alpha b(\mu_a-x)/v_a]', \\
\label{adot3}
\dot a &= D_a a'' + \beta_0 b + \beta_1g(x-\mu_b)b- \nu a.
\end{align}
Comparing to Eqs.\ \ref{bdot2} and \ref{adot2}, we see that Eq.\ \ref{bdot3} is unchanged, while Eq.\ \ref{adot3} (with $\beta\to\beta_0$) has an added term. Because Eq.\ \ref{bdot3} is unchanged, the equations for $\dot B$, $\dot \mu_b$, and $\dot v_b$ are unchanged.

Integrating Eq.\ \ref{adot3} gives a new term $\beta_1g\int dx\ (x-\mu_b)b$ on the righthand side, but this term vanishes by Eqs.\ \ref{mom1} and \ref{mom3}, leaving Eq.\ \ref{noc2A} for $\dot A$ unchanged. This invariance reflects the adaptive nature of secretion: the co-attractant is redistributed in space, but unchanged in amount.

Integrating Eq.\ \ref{adot3} against $x$ gives a new term $\beta_1g\int dx\ x(x-\mu_b)b$ on the righthand side, which evaluates to $\beta_1gBv_b$ using Eqs.\ \ref{mom3} and \ref{mom5} and simplifying. Thus, Eq.\ \ref{noc4A} becomes
\begin{equation}
\label{c4A}
\dot \mu_a = \beta_0 B(\mu_b-\mu_a)/A + \beta_1gBv_b/A.
\end{equation}

Integrating Eq.\ \ref{adot3} against $x^2$ gives a new term $\beta_1g\int dx\ x^2(x-\mu_b)b$ on the righthand side. Using Eq.\ \ref{mom5} and the third moment of a Gaussian, $\int dx\ x^3b = B(3\mu_bv_b+\mu_b^3)$, this term becomes $2\beta_1gB\mu_bv_b$ after simplification. Meanwhile, the new term in $\dot \mu_a$ contributes a term $2\beta_1gB\mu_av_b$ to the lefthand side (see Eq.\ \ref{noc6temp}). Together, these terms make Eq.\ \ref{noc6A}
\begin{align}
\dot v_a =&\ 2D_a + \beta_0 B[(\mu_b-\mu_a)^2+v_b-v_a]/A \nonumber \\
\label{c6A}
&+ 2\beta_1gB(\mu_b-\mu_a)v_b/A.
\end{align}

In summary, the only changes are Eqs.\ \ref{c4A} and \ref{c6A}, which become Eqs.\ \ref{c4} and \ref{c6} upon defining $V_0\equiv D_b\gamma g$, $\Delta\equiv\mu_b-\mu_a$, and $\eta \equiv \beta_1/(\gamma\beta_0)$.

\subsection{Higher dimensions (Eq.\ \ref{var})}

In $d$ spatial dimensions, Eqs.\ \ref{eq:KS1} and \ref{eq:KS} with $c=0$ read
\begin{align}
\label{db1}
0 &= D_b\vec{\nabla}\cdot[\vec{\nabla} b - \alpha b\vec{\nabla}(\ln a)], \\
\label{da1}
0 &= D_a\nabla^2 a + \beta b - \nu a
\end{align}
in steady state. Because $c=0$, there is spherical symmetry, meaning that $a$ and $b$ are only functions of the radial coordinate $r$. Therefore, equations \ref{db1} and \ref{da1} become
\begin{align}
\label{db2}
0 &= D_br^{1-d}[r^{d-1} b' - \alpha r^{d-1}b(\ln a)']', \\
\label{da2}
0 &= D_ar^{1-d} (r^{d-1}a')' + \beta b - \nu a,
\end{align}
where now we use prime for the derivative with respect to $r$.

By symmetry, the means of the bacterial and co-attractant distributions must be equal in steady state, and we define this common location as our origin: $\mu_b=\mu_a=0$. Then, the Gaussian approximation of Eq.\ \ref{Gaussian} in $d$ dimensions becomes
\begin{equation}
\label{Gd}
b = \frac{Be^{-r^2/(2v_b)}}{(2\pi v_b)^{d/2}}, \quad
a = \frac{Ae^{-r^2/(2v_a)}}{(2\pi v_a)^{d/2}},
\end{equation}
where again we use $\sigma_b^2\equiv v_b$ and $\sigma_a^2\equiv v_a$ for brevity. From Eq.\ \ref{Gd} we see that $(\ln a)' = -r/v_a$, making Eq.\ \ref{db2}
\begin{equation}
\label{db3}
0 = D_br^{1-d}(r^{d-1} b' + \alpha r^db/v_a)'.
\end{equation}

The integral over $d$-dimensional space $\int d^dx$ reduces to $\Omega\int dr\ r^{d-1}$ when the integrand is spherically symmetric, where $\Omega = 2\pi^{d/2}/\Gamma(d/2)$ is the angular component (i.e., $\Omega=2, 2\pi, 4\pi, \dots$ for $d=1,2,3,\dots$). Thus, with $\mu_b=\mu_a=0$, the remaining moments are defined
\begin{align}
\label{dmom1}
B &= \Omega\int_0^\infty dr\ r^{d-1}b, \\
\label{dmom2}
A &= \Omega\int_0^\infty dr\ r^{d-1}a, \\
\label{dmom3}
v_b &= \frac{\Omega}{Bd}\int_0^\infty dr\ r^{d-1}r^2b, \\
\label{dmom4}
v_a &= \frac{\Omega}{Ad}\int_0^\infty dr\ r^{d-1}r^2a.
\end{align}
In Eqs.\ \ref{dmom3} and \ref{dmom4}, the factor of $d$ in the denominator comes from the fact that the variance is the integral against the square of any of the Cartesian components, and we must sum all $d$ such squares to obtain the $r^2$.

Multiplying Eq.\ \ref{da2} against $\Omega r^{d-1}$ and integrating gives
\begin{equation}
0 = \Omega D_a\int_0^\infty dr\ (r^{d-1}a')' + \beta B - \nu A.
\end{equation}
Because the term in parentheses vanishes at the boundaries, the integral vanishes, leaving
\begin{equation}
\label{ss2}
0 = \beta B - \nu A,
\end{equation}
as we found in steady state when $d=1$ (Eq.\ \ref{noc2A}).

Multiplying Eq.\ \ref{db3} against $\Omega r^{d-1}r^2/(Bd)$ and integrating gives
\begin{equation}
0 = \frac{\Omega D_b}{Bd}\int_0^\infty dr\ r^2(r^{d-1} b' + \alpha r^db/v_a)'.
\end{equation}
Integrating by parts gives
\begin{equation}
0 = -\frac{2\Omega D_b}{Bd}\int_0^\infty dr\ r(r^{d-1} b' + \alpha r^db/v_a),
\end{equation}
since the boundary terms vanish as before. Integrating the first term by parts again gives
\begin{equation}
0 = \frac{2\Omega D_b}{Bd}\int_0^\infty dr\ (r^{d-1} bd - \alpha r^{d-1}r^2b/v_a).
\end{equation}
Using Eqs.\ \ref{dmom1} and \ref{dmom3}, this becomes
\begin{equation}
\label{ss5}
0 = 2D_b(1-\alpha v_b/v_a),
\end{equation}
as we found in steady state when $d=1$ (Eq.\ \ref{noc5A}).

Multiplying Eq.\ \ref{da2} against $\Omega r^{d-1}r^2/(Ad)$ and integrating gives
\begin{equation}
0 = \frac{\Omega D_a}{Ad} \int_0^\infty dr\ r^2(r^{d-1}a')' + \beta Bv_b/A - \nu v_a,
\end{equation}
where we used Eqs.\ \ref{dmom3} and \ref{dmom4}. As before, integrating the first term by parts twice makes it $2D_a$, giving
\begin{equation}
0 = 2D_a + \beta Bv_b/A - \nu v_a.
\end{equation}
Using Eq.\ \ref{ss2} to eliminate $\nu$, we have
\begin{equation}
\label{ss6}
0 = 2D_a + \beta B(v_b-v_a)/A,
\end{equation}
as we found with $\mu_b=\mu_a$ in steady state when $d=1$ (Eq.\ \ref{noc6A}).

Combining Eqs.\ \ref{ss2}, \ref{ss5}, and \ref{ss6} gives Eq.\ \ref{var}. Thus, we see that Eq.\ \ref{var} holds in any spatial dimension.

\section{Fast-communication approximation}
\label{fast}
Here we justify approximating Eqs.\ \ref{var2} and \ref{var3} by Eq.\ \ref{var} in the limit of fast communication. Approximating Eq.\ \ref{var2} by Eq.\ \ref{var} is valid if $D_a\gg\nu\Delta^2$. To find $\Delta$, we will assume that Eq.\ \ref{var} holds and then assess this condition post hoc.
Assuming that Eq.\ \ref{var} holds gives Eq.\ \ref{V} for $\Delta$, which we write as
\begin{equation}
\Delta = \frac{V_0}{\nu}\left[1+\frac{D_b}{D_a}\left(\frac{\alpha-1}{2}\right)\right]^{-1}.
\end{equation}
We see that $\Delta\sim V_0/\nu$, as long as $\alpha$ is not too large and $D_b$ is on the same order as $D_a$. {\it Escherichia coli} swim at tens of microns per second and tumble every second or so \cite{berg2025random}, making $D_b$ hundreds of square microns per second, which is indeed on the order of the diffusion coefficient of a small molecule in water. In particular, we use $D_b=D_a=400\ \mu$m$^2$/s in our numerical results.

Using $\Delta\sim V_0/\nu$, our condition $D_a\gg\nu\Delta^2$ becomes $D_a\nu\gg V_0^2$. This expression states that the diffusion ($D_a$) and turnover ($\nu$) of the communication molecule must be faster than the squared drift speed ($V_0^2$) of the cells. Hence we call this the fast communication limit. Typical {\it E.\ coli} chemotactic drift speeds are on the order of microns per second \cite{grognot2021multiscale}, which means that this condition will hold as long as the communication molecules degrade faster than about once per minute. This is a modest constraint, and therefore we expect this condition to hold. In particular, the parameters we use in our numerical results give $D_a\nu/V_0^2 = 62.5 \gg 1$.

Approximating Eq.\ \ref{var3} by Eq.\ \ref{var} is valid if additionally $\alpha-1\gg\eta V_0\Delta/D_b$. Assuming that $\alpha-1$ and $\eta$ are order one, and using $\Delta\sim V_0/\nu$, this condition becomes $D_b\nu\gg V_0^2$. Given $D_b\sim D_a$, we see that this condition is equivalent to the fast communication limit, $D_a\nu\gg V_0^2$.

\section{Positivity of secretion rate}
\label{positivity}
Noting from Eq.\ \ref{c} that $\ln c - \langle\ln c\rangle = g(x-\mu_b)$, and recalling the definition of $\eta$ (Eq.\ \ref{eta}), Eq.\ \ref{beta} reads
\begin{equation}
\label{beta2}
\beta = \beta_0[1+\gamma\eta g(x-\mu_b)].
\end{equation}
For sufficiently large $\eta$, a potential concern is that cells at the back of the bacterial distribution ($x<\mu_b$) could have a negative secretion rate ($\beta<0$). Here we show that even for $\eta>\eta_c$, the fast-communication limit of the previous appendix, $D_a\nu\gg V_0^2$, makes this unlikely to occur.

We require $\beta > 0$, even for cells a few standard deviations behind the mean, i.e., $x-\mu_b = -n\sigma_b$ (where $n$ is, say, 2 or 3). Inserting these two expressions into Eq.\ \ref{beta2} gives
\begin{equation}
\label{eta1}
\eta < \frac{1}{n\gamma g\sigma_b} = \frac{1}{n\gamma g}\sqrt{\frac{\nu(\alpha-1)}{2 D_a}},
\end{equation}
where the second step inserts Eq.\ \ref{var} for $\sigma_b$. For chemotaxis speedup, we need
\begin{equation}
\label{eta2}
\eta > \eta_c = \frac{(\alpha-1)D_b}{2D_a}
\end{equation}
from Eq.\ \ref{etac}. Equations \ref{eta1} and \ref{eta2} are compatible if
\begin{equation}
\frac{(\alpha-1)D_b}{2D_a} < \frac{1}{n\gamma g}\sqrt{\frac{\nu(\alpha-1)}{2 D_a}}.
\end{equation}
Recalling that $V_0\equiv D_b\gamma g$, this condition becomes
\begin{equation}
\frac{n^2(\alpha-1)}{2} < \frac{D_a\nu}{V_0^2}.
\end{equation}
In the fast-communication limit $D_a\nu\gg V_0^2$, the righthand side of this condition is much larger than one, making the condition easy to satisfy for modest values of $n$ and $\alpha$. In particular, speedup is most pronounced for $\alpha$ closest to 1 (Fig.\ \ref{fig:adaptive}), making satisfaction most likely.

%\bibliography{refs}

\begin{thebibliography}{38}%
\makeatletter
\providecommand \@ifxundefined [1]{%
 \@ifx{#1\undefined}
}%
\providecommand \@ifnum [1]{%
 \ifnum #1\expandafter \@firstoftwo
 \else \expandafter \@secondoftwo
 \fi
}%
\providecommand \@ifx [1]{%
 \ifx #1\expandafter \@firstoftwo
 \else \expandafter \@secondoftwo
 \fi
}%
\providecommand \natexlab [1]{#1}%
\providecommand \enquote  [1]{``#1''}%
\providecommand \bibnamefont  [1]{#1}%
\providecommand \bibfnamefont [1]{#1}%
\providecommand \citenamefont [1]{#1}%
\providecommand \href@noop [0]{\@secondoftwo}%
\providecommand \href [0]{\begingroup \@sanitize@url \@href}%
\providecommand \@href[1]{\@@startlink{#1}\@@href}%
\providecommand \@@href[1]{\endgroup#1\@@endlink}%
\providecommand \@sanitize@url [0]{\catcode `\\12\catcode `\$12\catcode
  `\&12\catcode `\#12\catcode `\^12\catcode `\_12\catcode `\%12\relax}%
\providecommand \@@startlink[1]{}%
\providecommand \@@endlink[0]{}%
\providecommand \url  [0]{\begingroup\@sanitize@url \@url }%
\providecommand \@url [1]{\endgroup\@href {#1}{\urlprefix }}%
\providecommand \urlprefix  [0]{URL }%
\providecommand \Eprint [0]{\href }%
\providecommand \doibase [0]{https://doi.org/}%
\providecommand \selectlanguage [0]{\@gobble}%
\providecommand \bibinfo  [0]{\@secondoftwo}%
\providecommand \bibfield  [0]{\@secondoftwo}%
\providecommand \translation [1]{[#1]}%
\providecommand \BibitemOpen [0]{}%
\providecommand \bibitemStop [0]{}%
\providecommand \bibitemNoStop [0]{.\EOS\space}%
\providecommand \EOS [0]{\spacefactor3000\relax}%
\providecommand \BibitemShut  [1]{\csname bibitem#1\endcsname}%
\let\auto@bib@innerbib\@empty
%</preamble>
\bibitem [{\citenamefont {Su}\ \emph {et~al.}(2024)\citenamefont {Su},
  \citenamefont {Song}, \citenamefont {Zhu}, \citenamefont {Huang},
  \citenamefont {Fan}, \citenamefont {Qiao},\ and\ \citenamefont
  {Mao}}]{su2024cell}%
  \BibitemOpen
  \bibfield  {author} {\bibinfo {author} {\bibfnamefont {J.}~\bibnamefont
  {Su}}, \bibinfo {author} {\bibfnamefont {Y.}~\bibnamefont {Song}}, \bibinfo
  {author} {\bibfnamefont {Z.}~\bibnamefont {Zhu}}, \bibinfo {author}
  {\bibfnamefont {X.}~\bibnamefont {Huang}}, \bibinfo {author} {\bibfnamefont
  {J.}~\bibnamefont {Fan}}, \bibinfo {author} {\bibfnamefont {J.}~\bibnamefont
  {Qiao}},\ and\ \bibinfo {author} {\bibfnamefont {F.}~\bibnamefont {Mao}},\
  }\bibfield  {title} {\bibinfo {title} {Cell--cell communication: new insights
  and clinical implications},\ }\href@noop {} {\bibfield  {journal} {\bibinfo
  {journal} {Signal Transduction and Targeted Therapy}\ }\textbf {\bibinfo
  {volume} {9}},\ \bibinfo {pages} {196} (\bibinfo {year} {2024})}\BibitemShut
  {NoStop}%
\bibitem [{\citenamefont {Fancher}\ and\ \citenamefont
  {Mugler}(2017)}]{fancher2017fundamental}%
  \BibitemOpen
  \bibfield  {author} {\bibinfo {author} {\bibfnamefont {S.}~\bibnamefont
  {Fancher}}\ and\ \bibinfo {author} {\bibfnamefont {A.}~\bibnamefont
  {Mugler}},\ }\bibfield  {title} {\bibinfo {title} {Fundamental limits to
  collective concentration sensing in cell populations},\ }\href@noop {}
  {\bibfield  {journal} {\bibinfo  {journal} {Physical review letters}\
  }\textbf {\bibinfo {volume} {118}},\ \bibinfo {pages} {078101} (\bibinfo
  {year} {2017})}\BibitemShut {NoStop}%
\bibitem [{\citenamefont {Mukherjee}\ and\ \citenamefont
  {Bassler}(2019)}]{mukherjee2019bacterial}%
  \BibitemOpen
  \bibfield  {author} {\bibinfo {author} {\bibfnamefont {S.}~\bibnamefont
  {Mukherjee}}\ and\ \bibinfo {author} {\bibfnamefont {B.~L.}\ \bibnamefont
  {Bassler}},\ }\bibfield  {title} {\bibinfo {title} {Bacterial quorum sensing
  in complex and dynamically changing environments},\ }\href@noop {} {\bibfield
   {journal} {\bibinfo  {journal} {Nature Reviews Microbiology}\ }\textbf
  {\bibinfo {volume} {17}},\ \bibinfo {pages} {371} (\bibinfo {year}
  {2019})}\BibitemShut {NoStop}%
\bibitem [{\citenamefont {Huang}\ \emph {et~al.}(2020)\citenamefont {Huang},
  \citenamefont {Chen},\ and\ \citenamefont {Zhang}}]{huang2020roles}%
  \BibitemOpen
  \bibfield  {author} {\bibinfo {author} {\bibfnamefont {Y.}~\bibnamefont
  {Huang}}, \bibinfo {author} {\bibfnamefont {Y.}~\bibnamefont {Chen}},\ and\
  \bibinfo {author} {\bibfnamefont {L.-h.}\ \bibnamefont {Zhang}},\ }\bibfield
  {title} {\bibinfo {title} {The roles of microbial cell-cell chemical
  communication systems in the modulation of antimicrobial resistance},\
  }\href@noop {} {\bibfield  {journal} {\bibinfo  {journal} {Antibiotics}\
  }\textbf {\bibinfo {volume} {9}},\ \bibinfo {pages} {779} (\bibinfo {year}
  {2020})}\BibitemShut {NoStop}%
\bibitem [{\citenamefont {Dominiak}\ \emph {et~al.}(2020)\citenamefont
  {Dominiak}, \citenamefont {Che{\l}stowska}, \citenamefont {Olejarz},\ and\
  \citenamefont {Nowicka}}]{dominiak2020communication}%
  \BibitemOpen
  \bibfield  {author} {\bibinfo {author} {\bibfnamefont {A.}~\bibnamefont
  {Dominiak}}, \bibinfo {author} {\bibfnamefont {B.}~\bibnamefont
  {Che{\l}stowska}}, \bibinfo {author} {\bibfnamefont {W.}~\bibnamefont
  {Olejarz}},\ and\ \bibinfo {author} {\bibfnamefont {G.}~\bibnamefont
  {Nowicka}},\ }\bibfield  {title} {\bibinfo {title} {Communication in the
  cancer microenvironment as a target for therapeutic interventions},\
  }\href@noop {} {\bibfield  {journal} {\bibinfo  {journal} {Cancers}\ }\textbf
  {\bibinfo {volume} {12}},\ \bibinfo {pages} {1232} (\bibinfo {year}
  {2020})}\BibitemShut {NoStop}%
\bibitem [{\citenamefont {Gilbert}(2014)}]{gilbert2014developmental}%
  \BibitemOpen
  \bibfield  {author} {\bibinfo {author} {\bibfnamefont {S.~F.}\ \bibnamefont
  {Gilbert}},\ }\href@noop {} {\bibinfo {title} {Developmental biology tenth
  edition}} (\bibinfo {year} {2014})\BibitemShut {NoStop}%
\bibitem [{\citenamefont {Aman}\ and\ \citenamefont
  {Piotrowski}(2010)}]{aman2010cell}%
  \BibitemOpen
  \bibfield  {author} {\bibinfo {author} {\bibfnamefont {A.}~\bibnamefont
  {Aman}}\ and\ \bibinfo {author} {\bibfnamefont {T.}~\bibnamefont
  {Piotrowski}},\ }\bibfield  {title} {\bibinfo {title} {Cell migration during
  morphogenesis},\ }\href@noop {} {\bibfield  {journal} {\bibinfo  {journal}
  {Developmental biology}\ }\textbf {\bibinfo {volume} {341}},\ \bibinfo
  {pages} {20} (\bibinfo {year} {2010})}\BibitemShut {NoStop}%
\bibitem [{\citenamefont {Cheung}\ and\ \citenamefont
  {Horne-Badovinac}(2025)}]{cheung2025collective}%
  \BibitemOpen
  \bibfield  {author} {\bibinfo {author} {\bibfnamefont {K.~J.}\ \bibnamefont
  {Cheung}}\ and\ \bibinfo {author} {\bibfnamefont {S.}~\bibnamefont
  {Horne-Badovinac}},\ }\bibfield  {title} {\bibinfo {title} {Collective
  migration modes in development, tissue repair and cancer},\ }\href@noop {}
  {\bibfield  {journal} {\bibinfo  {journal} {Nature Reviews Molecular Cell
  Biology}\ ,\ \bibinfo {pages} {1}} (\bibinfo {year} {2025})}\BibitemShut
  {NoStop}%
\bibitem [{\citenamefont {Luster}\ \emph {et~al.}(2005)\citenamefont {Luster},
  \citenamefont {Alon},\ and\ \citenamefont {Von~Andrian}}]{luster2005immune}%
  \BibitemOpen
  \bibfield  {author} {\bibinfo {author} {\bibfnamefont {A.~D.}\ \bibnamefont
  {Luster}}, \bibinfo {author} {\bibfnamefont {R.}~\bibnamefont {Alon}},\ and\
  \bibinfo {author} {\bibfnamefont {U.~H.}\ \bibnamefont {Von~Andrian}},\
  }\bibfield  {title} {\bibinfo {title} {Immune cell migration in inflammation:
  present and future therapeutic targets},\ }\href@noop {} {\bibfield
  {journal} {\bibinfo  {journal} {Nature immunology}\ }\textbf {\bibinfo
  {volume} {6}},\ \bibinfo {pages} {1182} (\bibinfo {year} {2005})}\BibitemShut
  {NoStop}%
\bibitem [{\citenamefont {Stuelten}\ \emph {et~al.}(2018)\citenamefont
  {Stuelten}, \citenamefont {Parent},\ and\ \citenamefont
  {Montell}}]{stuelten2018cell}%
  \BibitemOpen
  \bibfield  {author} {\bibinfo {author} {\bibfnamefont {C.~H.}\ \bibnamefont
  {Stuelten}}, \bibinfo {author} {\bibfnamefont {C.~A.}\ \bibnamefont
  {Parent}},\ and\ \bibinfo {author} {\bibfnamefont {D.~J.}\ \bibnamefont
  {Montell}},\ }\bibfield  {title} {\bibinfo {title} {Cell motility in cancer
  invasion and metastasis: insights from simple model organisms},\ }\href@noop
  {} {\bibfield  {journal} {\bibinfo  {journal} {Nature Reviews Cancer}\
  }\textbf {\bibinfo {volume} {18}},\ \bibinfo {pages} {296} (\bibinfo {year}
  {2018})}\BibitemShut {NoStop}%
\bibitem [{\citenamefont {Wadhwa}\ and\ \citenamefont
  {Berg}(2022)}]{wadhwa2022bacterial}%
  \BibitemOpen
  \bibfield  {author} {\bibinfo {author} {\bibfnamefont {N.}~\bibnamefont
  {Wadhwa}}\ and\ \bibinfo {author} {\bibfnamefont {H.~C.}\ \bibnamefont
  {Berg}},\ }\bibfield  {title} {\bibinfo {title} {Bacterial motility:
  machinery and mechanisms},\ }\href@noop {} {\bibfield  {journal} {\bibinfo
  {journal} {Nature reviews microbiology}\ }\textbf {\bibinfo {volume} {20}},\
  \bibinfo {pages} {161} (\bibinfo {year} {2022})}\BibitemShut {NoStop}%
\bibitem [{\citenamefont {Tu}(2013)}]{tu2013quantitative}%
  \BibitemOpen
  \bibfield  {author} {\bibinfo {author} {\bibfnamefont {Y.}~\bibnamefont
  {Tu}},\ }\bibfield  {title} {\bibinfo {title} {Quantitative modeling of
  bacterial chemotaxis: signal amplification and accurate adaptation},\
  }\href@noop {} {\bibfield  {journal} {\bibinfo  {journal} {Annual review of
  biophysics}\ }\textbf {\bibinfo {volume} {42}},\ \bibinfo {pages} {337}
  (\bibinfo {year} {2013})}\BibitemShut {NoStop}%
\bibitem [{\citenamefont {Jacob}\ \emph {et~al.}(2004)\citenamefont {Jacob},
  \citenamefont {Becker}, \citenamefont {Shapira},\ and\ \citenamefont
  {Levine}}]{jacob2004bacterial}%
  \BibitemOpen
  \bibfield  {author} {\bibinfo {author} {\bibfnamefont {E.~B.}\ \bibnamefont
  {Jacob}}, \bibinfo {author} {\bibfnamefont {I.}~\bibnamefont {Becker}},
  \bibinfo {author} {\bibfnamefont {Y.}~\bibnamefont {Shapira}},\ and\ \bibinfo
  {author} {\bibfnamefont {H.}~\bibnamefont {Levine}},\ }\bibfield  {title}
  {\bibinfo {title} {Bacterial linguistic communication and social
  intelligence},\ }\href@noop {} {\bibfield  {journal} {\bibinfo  {journal}
  {TRENDS in Microbiology}\ }\textbf {\bibinfo {volume} {12}},\ \bibinfo
  {pages} {366} (\bibinfo {year} {2004})}\BibitemShut {NoStop}%
\bibitem [{\citenamefont {Budrene}\ and\ \citenamefont
  {Berg}(1991)}]{budrene1991complex}%
  \BibitemOpen
  \bibfield  {author} {\bibinfo {author} {\bibfnamefont {E.~O.}\ \bibnamefont
  {Budrene}}\ and\ \bibinfo {author} {\bibfnamefont {H.~C.}\ \bibnamefont
  {Berg}},\ }\bibfield  {title} {\bibinfo {title} {Complex patterns formed by
  motile cells of escherichia coli},\ }\href@noop {} {\bibfield  {journal}
  {\bibinfo  {journal} {Nature}\ }\textbf {\bibinfo {volume} {349}},\ \bibinfo
  {pages} {630} (\bibinfo {year} {1991})}\BibitemShut {NoStop}%
\bibitem [{\citenamefont {Park}\ \emph
  {et~al.}(2003{\natexlab{a}})\citenamefont {Park}, \citenamefont {Wolanin},
  \citenamefont {Yuzbashyan}, \citenamefont {Silberzan}, \citenamefont
  {Stock},\ and\ \citenamefont {Austin}}]{park2003motion}%
  \BibitemOpen
  \bibfield  {author} {\bibinfo {author} {\bibfnamefont {S.}~\bibnamefont
  {Park}}, \bibinfo {author} {\bibfnamefont {P.~M.}\ \bibnamefont {Wolanin}},
  \bibinfo {author} {\bibfnamefont {E.~A.}\ \bibnamefont {Yuzbashyan}},
  \bibinfo {author} {\bibfnamefont {P.}~\bibnamefont {Silberzan}}, \bibinfo
  {author} {\bibfnamefont {J.~B.}\ \bibnamefont {Stock}},\ and\ \bibinfo
  {author} {\bibfnamefont {R.~H.}\ \bibnamefont {Austin}},\ }\bibfield  {title}
  {\bibinfo {title} {Motion to form a quorum},\ }\href@noop {} {\bibfield
  {journal} {\bibinfo  {journal} {Science}\ }\textbf {\bibinfo {volume}
  {301}},\ \bibinfo {pages} {188} (\bibinfo {year}
  {2003}{\natexlab{a}})}\BibitemShut {NoStop}%
\bibitem [{\citenamefont {Park}\ \emph
  {et~al.}(2003{\natexlab{b}})\citenamefont {Park}, \citenamefont {Wolanin},
  \citenamefont {Yuzbashyan}, \citenamefont {Lin}, \citenamefont {Darnton},
  \citenamefont {Stock}, \citenamefont {Silberzan},\ and\ \citenamefont
  {Austin}}]{park2003influence}%
  \BibitemOpen
  \bibfield  {author} {\bibinfo {author} {\bibfnamefont {S.}~\bibnamefont
  {Park}}, \bibinfo {author} {\bibfnamefont {P.~M.}\ \bibnamefont {Wolanin}},
  \bibinfo {author} {\bibfnamefont {E.~A.}\ \bibnamefont {Yuzbashyan}},
  \bibinfo {author} {\bibfnamefont {H.}~\bibnamefont {Lin}}, \bibinfo {author}
  {\bibfnamefont {N.~C.}\ \bibnamefont {Darnton}}, \bibinfo {author}
  {\bibfnamefont {J.~B.}\ \bibnamefont {Stock}}, \bibinfo {author}
  {\bibfnamefont {P.}~\bibnamefont {Silberzan}},\ and\ \bibinfo {author}
  {\bibfnamefont {R.}~\bibnamefont {Austin}},\ }\bibfield  {title} {\bibinfo
  {title} {Influence of topology on bacterial social interaction},\ }\href@noop
  {} {\bibfield  {journal} {\bibinfo  {journal} {Proceedings of the National
  Academy of Sciences}\ }\textbf {\bibinfo {volume} {100}},\ \bibinfo {pages}
  {13910} (\bibinfo {year} {2003}{\natexlab{b}})}\BibitemShut {NoStop}%
\bibitem [{\citenamefont {Long}\ \emph {et~al.}(2017)\citenamefont {Long},
  \citenamefont {Quaife}, \citenamefont {Salman},\ and\ \citenamefont
  {Oltvai}}]{long2017cell}%
  \BibitemOpen
  \bibfield  {author} {\bibinfo {author} {\bibfnamefont {Z.}~\bibnamefont
  {Long}}, \bibinfo {author} {\bibfnamefont {B.}~\bibnamefont {Quaife}},
  \bibinfo {author} {\bibfnamefont {H.}~\bibnamefont {Salman}},\ and\ \bibinfo
  {author} {\bibfnamefont {Z.~N.}\ \bibnamefont {Oltvai}},\ }\bibfield  {title}
  {\bibinfo {title} {Cell-cell communication enhances bacterial chemotaxis
  toward external attractants},\ }\href@noop {} {\bibfield  {journal} {\bibinfo
   {journal} {Scientific reports}\ }\textbf {\bibinfo {volume} {7}},\ \bibinfo
  {pages} {12855} (\bibinfo {year} {2017})}\BibitemShut {NoStop}%
\bibitem [{\citenamefont {Keller}\ and\ \citenamefont
  {Segel}(1971)}]{keller1971traveling}%
  \BibitemOpen
  \bibfield  {author} {\bibinfo {author} {\bibfnamefont {E.~F.}\ \bibnamefont
  {Keller}}\ and\ \bibinfo {author} {\bibfnamefont {L.~A.}\ \bibnamefont
  {Segel}},\ }\bibfield  {title} {\bibinfo {title} {Traveling bands of
  chemotactic bacteria: a theoretical analysis},\ }\href@noop {} {\bibfield
  {journal} {\bibinfo  {journal} {Journal of theoretical biology}\ }\textbf
  {\bibinfo {volume} {30}},\ \bibinfo {pages} {235} (\bibinfo {year}
  {1971})}\BibitemShut {NoStop}%
\bibitem [{\citenamefont {Keller}\ and\ \citenamefont
  {Segel}(1970)}]{keller1970initiation}%
  \BibitemOpen
  \bibfield  {author} {\bibinfo {author} {\bibfnamefont {E.~F.}\ \bibnamefont
  {Keller}}\ and\ \bibinfo {author} {\bibfnamefont {L.~A.}\ \bibnamefont
  {Segel}},\ }\bibfield  {title} {\bibinfo {title} {Initiation of slime mold
  aggregation viewed as an instability},\ }\href@noop {} {\bibfield  {journal}
  {\bibinfo  {journal} {Journal of theoretical biology}\ }\textbf {\bibinfo
  {volume} {26}},\ \bibinfo {pages} {399} (\bibinfo {year} {1970})}\BibitemShut
  {NoStop}%
\bibitem [{\citenamefont {Tindall}\ \emph {et~al.}(2008)\citenamefont
  {Tindall}, \citenamefont {Maini}, \citenamefont {Porter},\ and\ \citenamefont
  {Armitage}}]{tindall2008overview}%
  \BibitemOpen
  \bibfield  {author} {\bibinfo {author} {\bibfnamefont {M.~J.}\ \bibnamefont
  {Tindall}}, \bibinfo {author} {\bibfnamefont {P.~K.}\ \bibnamefont {Maini}},
  \bibinfo {author} {\bibfnamefont {S.~L.}\ \bibnamefont {Porter}},\ and\
  \bibinfo {author} {\bibfnamefont {J.~P.}\ \bibnamefont {Armitage}},\
  }\bibfield  {title} {\bibinfo {title} {Overview of mathematical approaches
  used to model bacterial chemotaxis ii: bacterial populations},\ }\href@noop
  {} {\bibfield  {journal} {\bibinfo  {journal} {Bulletin of mathematical
  biology}\ }\textbf {\bibinfo {volume} {70}},\ \bibinfo {pages} {1570}
  (\bibinfo {year} {2008})}\BibitemShut {NoStop}%
\bibitem [{\citenamefont {Salman}\ \emph {et~al.}(2006)\citenamefont {Salman},
  \citenamefont {Zilman}, \citenamefont {Loverdo}, \citenamefont {Jeffroy},\
  and\ \citenamefont {Libchaber}}]{salman2006solitary}%
  \BibitemOpen
  \bibfield  {author} {\bibinfo {author} {\bibfnamefont {H.}~\bibnamefont
  {Salman}}, \bibinfo {author} {\bibfnamefont {A.}~\bibnamefont {Zilman}},
  \bibinfo {author} {\bibfnamefont {C.}~\bibnamefont {Loverdo}}, \bibinfo
  {author} {\bibfnamefont {M.}~\bibnamefont {Jeffroy}},\ and\ \bibinfo {author}
  {\bibfnamefont {A.}~\bibnamefont {Libchaber}},\ }\bibfield  {title} {\bibinfo
  {title} {Solitary modes of bacterial culture in a temperature gradient},\
  }\href@noop {} {\bibfield  {journal} {\bibinfo  {journal} {Physical review
  letters}\ }\textbf {\bibinfo {volume} {97}},\ \bibinfo {pages} {118101}
  (\bibinfo {year} {2006})}\BibitemShut {NoStop}%
\bibitem [{\citenamefont {Lazova}\ \emph {et~al.}(2011)\citenamefont {Lazova},
  \citenamefont {Ahmed}, \citenamefont {Bellomo}, \citenamefont {Stocker},\
  and\ \citenamefont {Shimizu}}]{lazova2011response}%
  \BibitemOpen
  \bibfield  {author} {\bibinfo {author} {\bibfnamefont {M.~D.}\ \bibnamefont
  {Lazova}}, \bibinfo {author} {\bibfnamefont {T.}~\bibnamefont {Ahmed}},
  \bibinfo {author} {\bibfnamefont {D.}~\bibnamefont {Bellomo}}, \bibinfo
  {author} {\bibfnamefont {R.}~\bibnamefont {Stocker}},\ and\ \bibinfo {author}
  {\bibfnamefont {T.~S.}\ \bibnamefont {Shimizu}},\ }\bibfield  {title}
  {\bibinfo {title} {Response rescaling in bacterial chemotaxis},\ }\href@noop
  {} {\bibfield  {journal} {\bibinfo  {journal} {Proceedings of the National
  Academy of Sciences}\ }\textbf {\bibinfo {volume} {108}},\ \bibinfo {pages}
  {13870} (\bibinfo {year} {2011})}\BibitemShut {NoStop}%
\bibitem [{\citenamefont {Wartlick}\ \emph {et~al.}(2009)\citenamefont
  {Wartlick}, \citenamefont {Kicheva},\ and\ \citenamefont
  {Gonz{\'a}lez-Gait{\'a}n}}]{wartlick2009morphogen}%
  \BibitemOpen
  \bibfield  {author} {\bibinfo {author} {\bibfnamefont {O.}~\bibnamefont
  {Wartlick}}, \bibinfo {author} {\bibfnamefont {A.}~\bibnamefont {Kicheva}},\
  and\ \bibinfo {author} {\bibfnamefont {M.}~\bibnamefont
  {Gonz{\'a}lez-Gait{\'a}n}},\ }\bibfield  {title} {\bibinfo {title} {Morphogen
  gradient formation},\ }\href@noop {} {\bibfield  {journal} {\bibinfo
  {journal} {Cold Spring Harbor perspectives in biology}\ }\textbf {\bibinfo
  {volume} {1}},\ \bibinfo {pages} {a001255} (\bibinfo {year}
  {2009})}\BibitemShut {NoStop}%
\bibitem [{cod()}]{code}%
  \BibitemOpen
  \href@noop {} {}\bibinfo {note}
  {\url{https://doi.org/10.5281/zenodo.17868091}}\BibitemShut {NoStop}%
\bibitem [{\citenamefont {Young}\ \emph {et~al.}(2019)\citenamefont {Young},
  \citenamefont {Demir}, \citenamefont {Salman}, \citenamefont {Ermentrout},\
  and\ \citenamefont {Rubin}}]{young2019interactions}%
  \BibitemOpen
  \bibfield  {author} {\bibinfo {author} {\bibfnamefont {G.}~\bibnamefont
  {Young}}, \bibinfo {author} {\bibfnamefont {M.}~\bibnamefont {Demir}},
  \bibinfo {author} {\bibfnamefont {H.}~\bibnamefont {Salman}}, \bibinfo
  {author} {\bibfnamefont {G.~B.}\ \bibnamefont {Ermentrout}},\ and\ \bibinfo
  {author} {\bibfnamefont {J.~E.}\ \bibnamefont {Rubin}},\ }\bibfield  {title}
  {\bibinfo {title} {Interactions of solitary pulses of e. coli in a
  one-dimensional nutrient gradient},\ }\href@noop {} {\bibfield  {journal}
  {\bibinfo  {journal} {Physica D: Nonlinear Phenomena}\ }\textbf {\bibinfo
  {volume} {395}},\ \bibinfo {pages} {24} (\bibinfo {year} {2019})}\BibitemShut
  {NoStop}%
\bibitem [{\citenamefont {Tang}\ \emph {et~al.}(2014)\citenamefont {Tang},
  \citenamefont {Wang}, \citenamefont {Shi}, \citenamefont {Iglesias},
  \citenamefont {Devreotes},\ and\ \citenamefont
  {Huang}}]{tang2014evolutionarily}%
  \BibitemOpen
  \bibfield  {author} {\bibinfo {author} {\bibfnamefont {M.}~\bibnamefont
  {Tang}}, \bibinfo {author} {\bibfnamefont {M.}~\bibnamefont {Wang}}, \bibinfo
  {author} {\bibfnamefont {C.}~\bibnamefont {Shi}}, \bibinfo {author}
  {\bibfnamefont {P.~A.}\ \bibnamefont {Iglesias}}, \bibinfo {author}
  {\bibfnamefont {P.~N.}\ \bibnamefont {Devreotes}},\ and\ \bibinfo {author}
  {\bibfnamefont {C.-H.}\ \bibnamefont {Huang}},\ }\bibfield  {title} {\bibinfo
  {title} {Evolutionarily conserved coupling of adaptive and excitable networks
  mediates eukaryotic chemotaxis},\ }\href@noop {} {\bibfield  {journal}
  {\bibinfo  {journal} {Nature communications}\ }\textbf {\bibinfo {volume}
  {5}},\ \bibinfo {pages} {5175} (\bibinfo {year} {2014})}\BibitemShut
  {NoStop}%
\bibitem [{\citenamefont {Takeda}\ \emph {et~al.}(2012)\citenamefont {Takeda},
  \citenamefont {Shao}, \citenamefont {Adler}, \citenamefont {Charest},
  \citenamefont {Loomis}, \citenamefont {Levine}, \citenamefont {Groisman},
  \citenamefont {Rappel},\ and\ \citenamefont {Firtel}}]{takeda2012incoherent}%
  \BibitemOpen
  \bibfield  {author} {\bibinfo {author} {\bibfnamefont {K.}~\bibnamefont
  {Takeda}}, \bibinfo {author} {\bibfnamefont {D.}~\bibnamefont {Shao}},
  \bibinfo {author} {\bibfnamefont {M.}~\bibnamefont {Adler}}, \bibinfo
  {author} {\bibfnamefont {P.~G.}\ \bibnamefont {Charest}}, \bibinfo {author}
  {\bibfnamefont {W.~F.}\ \bibnamefont {Loomis}}, \bibinfo {author}
  {\bibfnamefont {H.}~\bibnamefont {Levine}}, \bibinfo {author} {\bibfnamefont
  {A.}~\bibnamefont {Groisman}}, \bibinfo {author} {\bibfnamefont {W.-J.}\
  \bibnamefont {Rappel}},\ and\ \bibinfo {author} {\bibfnamefont {R.~A.}\
  \bibnamefont {Firtel}},\ }\bibfield  {title} {\bibinfo {title} {Incoherent
  feedforward control governs adaptation of activated ras in a eukaryotic
  chemotaxis pathway},\ }\href@noop {} {\bibfield  {journal} {\bibinfo
  {journal} {Science signaling}\ }\textbf {\bibinfo {volume} {5}},\ \bibinfo
  {pages} {ra2} (\bibinfo {year} {2012})}\BibitemShut {NoStop}%
\bibitem [{\citenamefont {Mugler}\ \emph {et~al.}(2016)\citenamefont {Mugler},
  \citenamefont {Levchenko},\ and\ \citenamefont
  {Nemenman}}]{mugler2016limits}%
  \BibitemOpen
  \bibfield  {author} {\bibinfo {author} {\bibfnamefont {A.}~\bibnamefont
  {Mugler}}, \bibinfo {author} {\bibfnamefont {A.}~\bibnamefont {Levchenko}},\
  and\ \bibinfo {author} {\bibfnamefont {I.}~\bibnamefont {Nemenman}},\
  }\bibfield  {title} {\bibinfo {title} {Limits to the precision of gradient
  sensing with spatial communication and temporal integration},\ }\href@noop {}
  {\bibfield  {journal} {\bibinfo  {journal} {Proceedings of the National
  Academy of Sciences}\ }\textbf {\bibinfo {volume} {113}},\ \bibinfo {pages}
  {E689} (\bibinfo {year} {2016})}\BibitemShut {NoStop}%
\bibitem [{\citenamefont {Camley}\ \emph {et~al.}(2016)\citenamefont {Camley},
  \citenamefont {Zimmermann}, \citenamefont {Levine},\ and\ \citenamefont
  {Rappel}}]{camley2016collective}%
  \BibitemOpen
  \bibfield  {author} {\bibinfo {author} {\bibfnamefont {B.~A.}\ \bibnamefont
  {Camley}}, \bibinfo {author} {\bibfnamefont {J.}~\bibnamefont {Zimmermann}},
  \bibinfo {author} {\bibfnamefont {H.}~\bibnamefont {Levine}},\ and\ \bibinfo
  {author} {\bibfnamefont {W.-J.}\ \bibnamefont {Rappel}},\ }\bibfield  {title}
  {\bibinfo {title} {Collective signal processing in cluster chemotaxis: Roles
  of adaptation, amplification, and co-attraction in collective guidance},\
  }\href@noop {} {\bibfield  {journal} {\bibinfo  {journal} {PLoS computational
  biology}\ }\textbf {\bibinfo {volume} {12}},\ \bibinfo {pages} {e1005008}
  (\bibinfo {year} {2016})}\BibitemShut {NoStop}%
\bibitem [{\citenamefont {Ellison}\ \emph {et~al.}(2016)\citenamefont
  {Ellison}, \citenamefont {Mugler}, \citenamefont {Brennan}, \citenamefont
  {Lee}, \citenamefont {Huebner}, \citenamefont {Shamir}, \citenamefont {Woo},
  \citenamefont {Kim}, \citenamefont {Amar}, \citenamefont {Nemenman} \emph
  {et~al.}}]{ellison2016cell}%
  \BibitemOpen
  \bibfield  {author} {\bibinfo {author} {\bibfnamefont {D.}~\bibnamefont
  {Ellison}}, \bibinfo {author} {\bibfnamefont {A.}~\bibnamefont {Mugler}},
  \bibinfo {author} {\bibfnamefont {M.~D.}\ \bibnamefont {Brennan}}, \bibinfo
  {author} {\bibfnamefont {S.~H.}\ \bibnamefont {Lee}}, \bibinfo {author}
  {\bibfnamefont {R.~J.}\ \bibnamefont {Huebner}}, \bibinfo {author}
  {\bibfnamefont {E.~R.}\ \bibnamefont {Shamir}}, \bibinfo {author}
  {\bibfnamefont {L.~A.}\ \bibnamefont {Woo}}, \bibinfo {author} {\bibfnamefont
  {J.}~\bibnamefont {Kim}}, \bibinfo {author} {\bibfnamefont {P.}~\bibnamefont
  {Amar}}, \bibinfo {author} {\bibfnamefont {I.}~\bibnamefont {Nemenman}},
  \emph {et~al.},\ }\bibfield  {title} {\bibinfo {title} {Cell--cell
  communication enhances the capacity of cell ensembles to sense shallow
  gradients during morphogenesis},\ }\href@noop {} {\bibfield  {journal}
  {\bibinfo  {journal} {Proceedings of the National Academy of Sciences}\
  }\textbf {\bibinfo {volume} {113}},\ \bibinfo {pages} {E679} (\bibinfo {year}
  {2016})}\BibitemShut {NoStop}%
\bibitem [{\citenamefont {Gonz{\'a}lez}\ and\ \citenamefont
  {Mugler}(2023)}]{gonzalez2023collective}%
  \BibitemOpen
  \bibfield  {author} {\bibinfo {author} {\bibfnamefont {L.}~\bibnamefont
  {Gonz{\'a}lez}}\ and\ \bibinfo {author} {\bibfnamefont {A.}~\bibnamefont
  {Mugler}},\ }\bibfield  {title} {\bibinfo {title} {Collective effects in
  flow-driven cell migration},\ }\href@noop {} {\bibfield  {journal} {\bibinfo
  {journal} {Physical Review E}\ }\textbf {\bibinfo {volume} {108}},\ \bibinfo
  {pages} {054406} (\bibinfo {year} {2023})}\BibitemShut {NoStop}%
\bibitem [{\citenamefont {Adler}(1966)}]{adler1966chemotaxis}%
  \BibitemOpen
  \bibfield  {author} {\bibinfo {author} {\bibfnamefont {J.}~\bibnamefont
  {Adler}},\ }\bibfield  {title} {\bibinfo {title} {Chemotaxis in bacteria:
  Motile escherichia coli migrate in bands that are influenced by oxygen and
  organic nutrients.},\ }\href@noop {} {\bibfield  {journal} {\bibinfo
  {journal} {Science}\ }\textbf {\bibinfo {volume} {153}},\ \bibinfo {pages}
  {708} (\bibinfo {year} {1966})}\BibitemShut {NoStop}%
\bibitem [{\citenamefont {Phan}\ \emph {et~al.}(2024)\citenamefont {Phan},
  \citenamefont {Mattingly}, \citenamefont {Vo}, \citenamefont {Marvin},
  \citenamefont {Looger},\ and\ \citenamefont {Emonet}}]{phan2024direct}%
  \BibitemOpen
  \bibfield  {author} {\bibinfo {author} {\bibfnamefont {T.~V.}\ \bibnamefont
  {Phan}}, \bibinfo {author} {\bibfnamefont {H.~H.}\ \bibnamefont {Mattingly}},
  \bibinfo {author} {\bibfnamefont {L.}~\bibnamefont {Vo}}, \bibinfo {author}
  {\bibfnamefont {J.~S.}\ \bibnamefont {Marvin}}, \bibinfo {author}
  {\bibfnamefont {L.~L.}\ \bibnamefont {Looger}},\ and\ \bibinfo {author}
  {\bibfnamefont {T.}~\bibnamefont {Emonet}},\ }\bibfield  {title} {\bibinfo
  {title} {Direct measurement of dynamic attractant gradients reveals breakdown
  of the patlak--keller--segel chemotaxis model},\ }\href@noop {} {\bibfield
  {journal} {\bibinfo  {journal} {Proceedings of the National Academy of
  Sciences}\ }\textbf {\bibinfo {volume} {121}},\ \bibinfo {pages}
  {e2309251121} (\bibinfo {year} {2024})}\BibitemShut {NoStop}%
\bibitem [{\citenamefont {Narla}\ \emph {et~al.}(2021)\citenamefont {Narla},
  \citenamefont {Cremer},\ and\ \citenamefont {Hwa}}]{narla2021traveling}%
  \BibitemOpen
  \bibfield  {author} {\bibinfo {author} {\bibfnamefont {A.~V.}\ \bibnamefont
  {Narla}}, \bibinfo {author} {\bibfnamefont {J.}~\bibnamefont {Cremer}},\ and\
  \bibinfo {author} {\bibfnamefont {T.}~\bibnamefont {Hwa}},\ }\bibfield
  {title} {\bibinfo {title} {A traveling-wave solution for bacterial chemotaxis
  with growth},\ }\href@noop {} {\bibfield  {journal} {\bibinfo  {journal}
  {Proceedings of the National Academy of Sciences}\ }\textbf {\bibinfo
  {volume} {118}},\ \bibinfo {pages} {e2105138118} (\bibinfo {year}
  {2021})}\BibitemShut {NoStop}%
\bibitem [{\citenamefont {Waite}\ \emph {et~al.}(2018)\citenamefont {Waite},
  \citenamefont {Frankel},\ and\ \citenamefont {Emonet}}]{waite2018behavioral}%
  \BibitemOpen
  \bibfield  {author} {\bibinfo {author} {\bibfnamefont {A.~J.}\ \bibnamefont
  {Waite}}, \bibinfo {author} {\bibfnamefont {N.~W.}\ \bibnamefont {Frankel}},\
  and\ \bibinfo {author} {\bibfnamefont {T.}~\bibnamefont {Emonet}},\
  }\bibfield  {title} {\bibinfo {title} {Behavioral variability and phenotypic
  diversity in bacterial chemotaxis},\ }\href@noop {} {\bibfield  {journal}
  {\bibinfo  {journal} {Annual review of biophysics}\ }\textbf {\bibinfo
  {volume} {47}},\ \bibinfo {pages} {595} (\bibinfo {year} {2018})}\BibitemShut
  {NoStop}%
\bibitem [{\citenamefont {Camley}(2018)}]{camley2018collective}%
  \BibitemOpen
  \bibfield  {author} {\bibinfo {author} {\bibfnamefont {B.~A.}\ \bibnamefont
  {Camley}},\ }\bibfield  {title} {\bibinfo {title} {Collective gradient
  sensing and chemotaxis: modeling and recent developments},\ }\href@noop {}
  {\bibfield  {journal} {\bibinfo  {journal} {Journal of Physics: Condensed
  Matter}\ }\textbf {\bibinfo {volume} {30}},\ \bibinfo {pages} {223001}
  (\bibinfo {year} {2018})}\BibitemShut {NoStop}%
\bibitem [{\citenamefont {Berg}(2025)}]{berg2025random}%
  \BibitemOpen
  \bibfield  {author} {\bibinfo {author} {\bibfnamefont {H.~C.}\ \bibnamefont
  {Berg}},\ }\href@noop {} {\emph {\bibinfo {title} {Random walks in
  biology}}}\ (\bibinfo  {publisher} {Princeton University Press},\ \bibinfo
  {year} {2025})\BibitemShut {NoStop}%
\bibitem [{\citenamefont {Grognot}\ and\ \citenamefont
  {Taute}(2021)}]{grognot2021multiscale}%
  \BibitemOpen
  \bibfield  {author} {\bibinfo {author} {\bibfnamefont {M.}~\bibnamefont
  {Grognot}}\ and\ \bibinfo {author} {\bibfnamefont {K.~M.}\ \bibnamefont
  {Taute}},\ }\bibfield  {title} {\bibinfo {title} {A multiscale 3d chemotaxis
  assay reveals bacterial navigation mechanisms},\ }\href@noop {} {\bibfield
  {journal} {\bibinfo  {journal} {Communications biology}\ }\textbf {\bibinfo
  {volume} {4}},\ \bibinfo {pages} {669} (\bibinfo {year} {2021})}\BibitemShut
  {NoStop}%
\end{thebibliography}

%apsrev4-2.bst 2019-01-14 (MD) hand-edited version of apsrev4-1.bst
%Control: key (0)
%Control: author (8) initials jnrlst
%Control: editor formatted (1) identically to author
%Control: production of article title (0) allowed
%Control: page (0) single
%Control: year (1) truncated
%Control: production of eprint (0) enabled
\providecommand{\noopsort}[1]{}\providecommand{\singleletter}[1]{#1}%

\end{document}